\documentclass[aps,prb,10pt,twocolumn,showpacs,
amsmath,groupedaddress]{revtex4}

\usepackage{epsfig}

\renewcommand{\vec}[1]{{\rm\bf #1}}
\renewcommand{\Re}{\mathop{\mathrm{Re}}\nolimits}
\renewcommand{\Im}{\mathop{\mathrm{Im}}\nolimits}

\newcommand{\ep}{\epsilon}

\begin{document}

\title{Disordered Josephson junction chains:
Anderson localization of normal modes and impedance fluctuations}

\author{D. M. Basko}
\affiliation{Universit\'e Grenoble 1/CNRS, LPMMC (UMR 5493), 25 rue des Martyrs, B.P.
166, 38042 Grenoble, France}
\author{F. W. J. Hekking}
\affiliation{Universit\'e Grenoble 1/CNRS, LPMMC (UMR 5493), 25 rue des Martyrs, B.P.
166, 38042 Grenoble, France}

\begin{abstract}
We study the properties of the normal modes of a chain of Josephson junctions in the simultaneous presence of disorder and absorption. We consider the superconducting regime of small phase fluctuations and focus on the case where the effects of disorder and absorption can be treated additively. We analyze the frequency shift and the localization length of the modes. We also calculate the distribution of the frequency-dependent impedance of the chain. The distribution is Gaussian if the localization length is long compared to the absorption length; it has a power law tail in the opposite limit.
\end{abstract}


\maketitle

\section{Introduction}

For more than a decade now, one-dimensional chains of Josephson junctions have been used
as controlled electromagnetic environments in experiments on superconducting
nanocircuits. This includes their use as high-impedance
environments~\cite{Watanabe2001,Corlevi2006a,Corlevi2006b}, and more recently as
superinductors~\cite{Manucharyan2009,Masluk2012,Bell2012}. Indeed, depending on the ratio
of the characteristic charging energy $E_C$ and Josephson coupling energy $E_J$,
Josephson junction chains can be tuned into the insulating regime, $E_C/E_J \gg 1$,
characterized by a highly resistive response, or the superconducting regime, $E_C/E_J
<1$, with a response dominated by the total effective Josephson
inductance~\cite{Bradley1984,Chow1998,Haviland2000,Kuo2001,Miyazaki2002,Takahide2006,Fazio2001}.
The use of chains made out of SQUID loops makes it possible to tune the ratio $E_C/E_J$
{\em in situ} experimentally by varying the applied magnetic flux~\cite{Haviland2000}.

In the superconducting regime, the fluctuations of the superconducting phase difference
across each junction in the chain are small~\cite{Fazio2001}. The chain behaves as an
effective LC transmission line, sustaining propagating electromagnetic modes.  Details of
the chain's electromagnetic response depend on the properties of these modes. In this
paper we focus on the superconducting regime and consider disordered chains, for which
the values of the parameters of the junctions forming the chain vary randomly from one
junction to the other. Dyson~\cite{Dyson1953} was the first to analyze the frequency
distribution of the normal modes of disordered LC transmission lines. Later, the spectral
properties of random chains were investigated in the framework of localization
phenomena~\cite{Ziman1982}. These studies neglected effects related to absorption.
Absorption should be taken into account in the case of Josephson junction chains, as
Josephson junctions are generally characterized by a finite quality
factor~\cite{Likharev1986,Schoen1990}. Effects of absorption and disorder have been
studied in Refs.~\onlinecite{Freilikher1994,PradhanKumar,BruceChalker,Beenakker1996,
Misirpashaev1996,Deych2001} for the one-dimensional Helmholtz equation with spatially
fluctuating dielectric constant.

In this paper, we analyze the effects of the simultaneous presence of disorder and
absorption on the electromagnetic properties of Josephson junction chains in the
superconducting regime. Here we consider the whole range of frequencies, and only in the
low-frequency limit can the Josephson junction chain be effectively described by the
Helmholtz equation. Specifically, we study the properties of the normal modes and
calculate the localization length and frequency shift for the case where absorption and
disorder act additively. We also study the statistics of the chain's frequency-dependent
impedance and calculate its distribution. The corresponding main results of the paper are
represented by Eqs.~(\ref{xi=}), (\ref{dwk2av=}), (\ref{PZ=}), and (\ref{PZtail=}).

Our results are relevant in view of the aforementioned experiments, in which uniform
Josephson junction chains are implemented as tunable environments in quantum
circuitry\cite{Watanabe2001,Corlevi2006a,Corlevi2006b,Manucharyan2009,Masluk2012,Bell2012}.
Disorder is inevitably present when fabricating Josephson junction arrays and knowledge
as to what disorder levels are acceptable in order for the chains to be uniform enough
for applications is important. On the other hand, our results can also be useful in the
context of mode engineering. Indeed, using intentionally induced disorder, certain modes
will be localized, thus suppressing the electromagnetic response of the chain at the
corresponding frequencies, which might be of interest for applications of chains as
electromagnetic environments.

The paper is organized as follows. We start by exposing the model in Section II and
present a qualitative discussion of the main results in Section III. The detailed
calculations are presented in Sections IV -- VI. Determining the normal modes for a
disordered Josephson junction chain in the presence of absorption corresponds to solving
a non-Hermitian eigenvalue problem which we address in section IV. The localization
length of the normal modes is calculated in Section V. The statistics of the chain's
impedance is analyzed in Section VI; Section VII contains our conclusions.

\section{The model}
\label{sec:Model}

\begin{figure}
\begin{center}
\vspace{0cm}
\includegraphics[width=8cm]{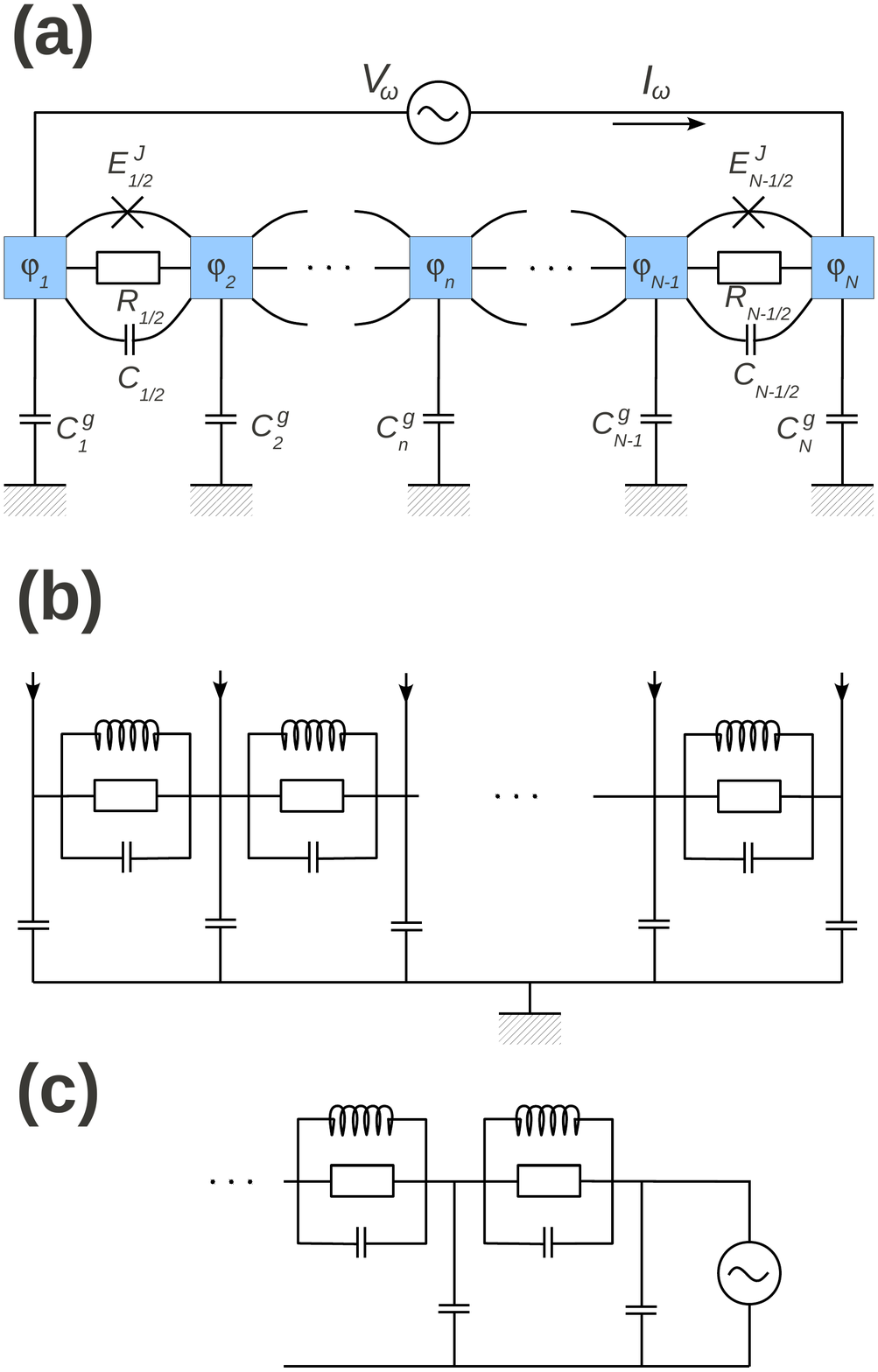}
\vspace{-0cm}
\end{center}
\caption{\label{fig:JJchain} (color online)
(a)~A schematic view of the Josephson junction chain and its
impedance measurement.
(b)~The transmission line described by Eqs.~(\ref{linsys=}),
equivalent to the chain shown in~(a) for $I_N=-I_1=I_\omega$,
$I_2,\ldots,I_{N-1}=0$, $V_N-V_1=V_\omega$.
(c)~The circuit used to define the impedance of a semi-infinite chain.}
\end{figure}

The chain to be studied in the present paper is assumed to
consist of $N$ superconducting islands, labelled by an integer
$n=1,\ldots,N$. The Josephson junction between the islands
$n$ and $n+1$ is labelled by the half-integer $n+1/2$. We focus
on small oscillations of the superconducting phase~$\theta_n$
of each island, so that the Josephson current through the
$(n+1/2)$th junction, $I^c_{n+1/2}\sin(\theta_{n+1}-\theta_n)$,
can be linearized as $I^c_{n+1/2}(\theta_{n+1}-\theta_n)$
here $I^c_{n+1/2}$ is the critical current of the junction,
related to the Josephson energy
$E^J_{n+1/2}=\hbar{}I^c_{n+1/2}/(2e)$].
The observable quantity on which we focus in the present paper
is the complex impedance $Z(\omega)$ of the chain at the
frequency~$\omega$, defined as the ratio of the voltage
$V_\omega$ on an external ac voltage source, connected to the
first and the last islands of the chain, to the current
$I_\omega$ through this source, as shown in
Fig.~\ref{fig:JJchain}(a). The complex impedance is in one-to-one
correspondence with the reflection coefficient of the equivalent
transmission line, shown in Fig.~\ref{fig:JJchain}(b), as discussed
in Appendix~\ref{app:reflection}.

For small oscillations, the superconducting phases of the
chain of $N$ islands at frequency~$\omega$ can be represented
as
\begin{equation}\label{thetan=}
\theta_n(t)=\theta_n(0)+\frac{2eV_n}\hbar\,
\frac{e^{-i\omega{t}}}{-i\omega},
\end{equation}
where $V_n$ is the ac voltage on the $n$th island, and $e<0$
is the electron charge. The voltages satisfy the following
system of linear equations:
\begin{widetext}\begin{equation}\begin{split}
&Y_{3/2}(V_1-V_2)-i\omega{C}^g_1V_1=I_1,\\
&Y_{n-1/2}(V_n-V_{n-1})+Y_{n+1/2}(V_n-V_{n+1})-i\omega{C}_n^gV_n
=I_n\quad (1<n<N),\\
&Y_{N-1/2}(V_N-V_{N-1})-i\omega{C}^g_NV_N=I_N.
\label{linsys=}
\end{split}\end{equation}\end{widetext}
The $n$th equation of this system is nothing but the first
Kirchhoff law (current conservation condition) at the $n$th
island. $I_n$~represents an external current injected into
the $n$th island (actually, the setting shown in
Fig.~\ref{fig:JJchain}(a) corresponds to only $I_N=-I_1$
being non-zero, but we include all $I_n$'s to formally
display the right-hand side of the linear system).
Each island is assumed to have some capacitance $C_n^g$ with
respect to the ground, so $-i\omega{C}_n^gV_n$ is the
displacement current leaking to the ground through this
capacitance.
$Y_{n+1/2}$~is the admittance of the $(n+1/2)$th junction,
so $Y_{n+1/2}(V_n-V_{n+1})$ is the current leaving the $n$th
island through this junction. In accordance with the standard RCSJ-model for Josephson junctions~\cite{Likharev1986,Schoen1990}, the admittance includes three
terms:
\begin{equation}
Y_{n+1/2}(\omega)=
-\frac{1}{i\omega{L}_{n+1/2}}-i\omega{C}_{n+1/2}
+\frac{1}{R_{n+1/2}}.
\end{equation}
The first one represents the linearized Josephson contribution
$I^c_{n+1/2}(\theta_{n+1}-\theta_n)$, by virtue of
Eq.~(\ref{thetan=}) and by the definition of the Josephson
inductance $L_{n+1/2}=-\hbar/(2eI^c_{n+1/2})$.
The second term is the contribution of the capacitive
electrostatic coupling between the neighboring islands.
Finally, $1/R_{n+1/2}$ is the dissipative junction conductance
due to the normal current carried by quasiparticles. It is
expected to vanish ($R_{n+1/2}\to\infty$) at zero temperature,
and approach the normal state conductance as the critical
temperature is approached.
The system~(\ref{linsys=}) corresponds to the effective electric
circuit shown in Fig.~\ref{fig:JJchain}(b).

For a weakly disordered Josephson junction chain, the
capacitances, inductances and resistances of its elements
can be represented as
\begin{equation}\begin{split}\label{disorder=}
&C^g_n=C^g(1+\eta_n),\\
&L_{n+1/2}=\frac{L}{1+\zeta_n},\\
&C_{n+1/2}=C(1+\zeta_n),
\end{split}\end{equation}
where the weak relative fluctuations
$\eta_n,\zeta_n$ are independent Gaussian random
variables with zero average and
\begin{equation}
\langle\eta_n^2\rangle=\sigma_g^2,\quad
\langle\zeta_n^2\rangle=\sigma_S^2.
\end{equation}
Here the angular brackets denote the statistical average, and
variables with different $n$'s are uncorrelated. The assumption
of weak disorder implies $\sigma_g,\sigma_S\ll{1}$. Note that the fluctuations of $L_{n+1/2}$ and $C_{n+1/2}$ are not independent: the product $L_{n+1/2} C_{n+1/2} \equiv 1/\omega_p^2$ is assumed to be constant, equal to the inverse squared junction plasma frequency $\omega_p$. This is because we assume that the fluctuations of $L$ and $C$ are due to fluctuations of the junction sizes $S$. Typically, $L \sim 1/S$, whereas $C\sim S$.

We do not consider the fluctuations of the normal resistances,
assuming $R_{n+1/2}=R$.
We are interested in the regime of large~$R$, when the average
effect of the resistance (namely, the absorption) is small;
weak fluctuations of~$R$ acting on top of this small average
have a negligible effect on the statistics of impedance, as
compared to the fluctuations of inductances and capacitances.
This can be checked directly by repeating the calculations of
Sec.~\ref{sec:Impedance} in the presence of fluctuations of~$R$;
they are fully analogous but more cumbersome, and the result is
quite trivial.
Thus, we prefer to neglect the fluctuations of~$R$ from the very
beginning.

\section{Qualitative discussion and summary of the main results}
\label{sec:Discussion}

We start by noting that the system~(\ref{linsys=}) can be
used to study two, generally speaking, physically distinct
problems.

\begin{figure}

\begin{center}
\vspace{0cm}
\includegraphics[width=8cm]{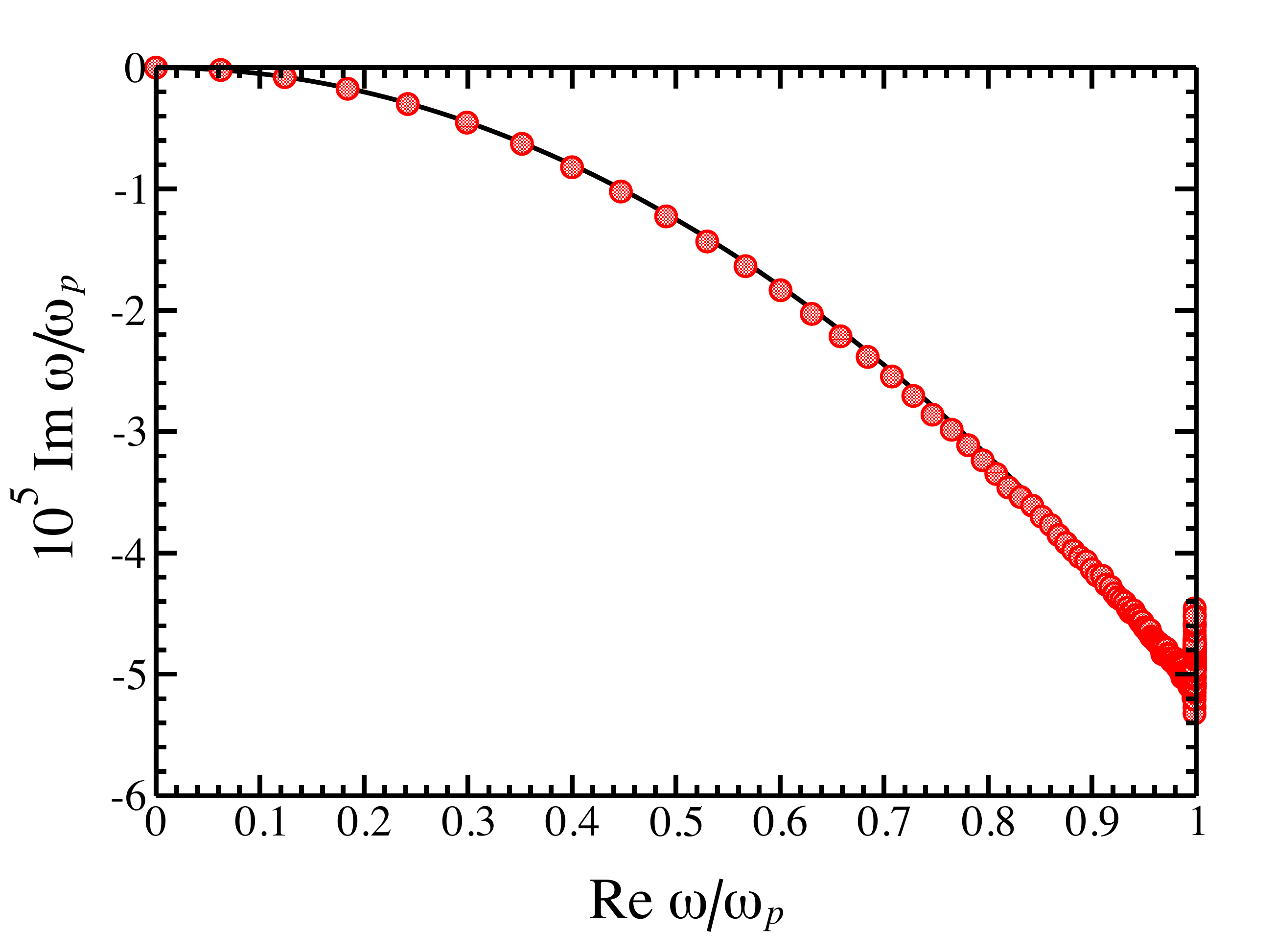}
\vspace{-0cm}
\end{center}
\caption{\label{fig:dispersion}
Complex plane of~$\omega$ in the units of $\omega_p\equiv(LC)^{-1/2}$.
The solid line shows the dispersion curve $\Omega(k)$ for
$\omega_pRC=10^4$, $C^g/C=0.01$.
The symbols represent the frequencies of the normal modes of a
disordered chain with $N=500$ islands and
$\sigma_S^2=0.1$, $\sigma_g^2=0$.
The condition $\xi(\omega)<N$ is fulfilled only for
$\omega/\omega_p>0.969\ldots$, so most of the interval
$0<\omega<\omega_p$ is occupied by modes which are weakly perturbed
by the disorder.
However, most of the modes (about 87\%) lie in the narrow frequency
interval with $\xi(\omega)<N$.
Strong fluctuations of $\Im\omega$, seen as the vertical feature on
the right edge of the figure, occur for the modes whose frequencies
lie close to the band edge of the clean chain. Their localization
lengh is $\xi\sim{1}$, so the approach used in the present paper is
not valid for their description.
}
\end{figure}

The first one is the problem of free oscillations (eigenmodes),
which consists in finding nontrivial solutions for the
voltages~$V_n$ when all external currents $I_n=0$. Such solutions
exist only for some special values of~$\omega$, which are,
generally speaking, complex, because of the dissipation induced
by the resistors. Physically, these solutions represent charge
distributions which oscillate and relax exponentially in
time while maintaining their spatial shape (which corresponds
to a gedanken experiment, rather than a real one). Mathematically,
this corresponds to a non-Hermitian quadratic eigenvalue problem,
to be discussed in detail in Sec.~\ref{sec:Eigen}.
In an infinite disorder-free chain the solutions are necessarily
plane waves, $V_n\propto{e}^{ikn}$, so the corresponding
frequencies define the dispersion relation $\Omega(k)$ which is
a complex function of a real argument~$k$. This dispersion relation
can be represented as a curve in the complex plane of~$\omega$
(Fig.~\ref{fig:dispersion}).
In a disordered system, the eigenmodes are no longer plane waves,
but are exponentially localized with some localization
length~$\xi$. Strictly speaking, their frequencies do not form
a continuous curve in the complex plane of $\omega$, but rather
represent a set of points. Still, when the disorder is weak, so
that the localization length $\xi$ is sufficiently large, the
uncertainty in the wave vector $1/\xi\ll{k}$, so the points lie
in the vicinity of the original dispersion curve of the
disorder-free chain (Fig.~\ref{fig:dispersion}). To the leading
order in the disorder strength, one can speak about the
localization length~$\xi$ as a function of the wave vector~$k$.

The second problem is that of forced oscillations, which consists
in finding the voltage profile $V_n$ in the presence of external
currents $I_n$, oscillating at a \emph{real} frequency~$\omega$.
In particular, the impedance of the chain~$Z(\omega)$, introduced
in the beginning of the previous section (Fig.~\ref{fig:JJchain}),
is found from the solution of such a problem with the currents
applied to the two ends of the chain, while away from the ends
$I_2=\ldots=I_{N-1}=0$. Mathematically, the problem is just to
invert the matrix of the system~(\ref{linsys=}).
In the absence of disorder, the voltage profile away from the
ends is still represented by plane waves.
However, in the presence of dissipation, the corresponding wave
vector must be complex: $V_n\propto{e}^{i(k+i\kappa)n}$, and
$k+i\kappa$ are determined by the solutions of the equation
$\Omega(k+i\kappa)=\omega$ with real~$\omega$. Physically,
this means that the ac excitation penetrates the chain only
within the distance $\sim{1}/\kappa$ from the ends, and at
longer distances it decays because of the absorption. Thus,
$1/\kappa$ can be called the absorption length. In a disordered
chain, the solution for $V_n$ decays away from the ends even in
the absence of dissipation, due to the localization, and the
corresponding length scale is the localization length~$\xi$.

When both dissipation and disorder are present, they both
contribute to the spatial decay of the solution (whose rate
is called Lyapunov exponent), and, generally speaking,
their effects are not easy to separate. For this reason, the
notion of the localization length as a function of frequency
in the complex plane is not very well defined in the presence
of dissipation. Still, in the limit of weak disorder and weak
dissipation the effects of localization and absorption can be
assumed to be additive. Namely, the Lyapunov exponent is given
by the sum $\kappa+1/\xi$ where $\kappa$ is calculated for weak
dissipation and no disorder (that is, small $1/R$ and
$\sigma_g^2,\sigma_S^2=0$)
while $\xi$ is calculated for weak disorder and no dissipation
(that is, $1/R=0$ and small $\sigma_g^2,\sigma_S^2$). Indeed,
the additive expression $\kappa+1/\xi$ is nothing but the first
(linear) term in the expansion of the Lyapunov exponent in the
small parameters $1/R,\sigma_g^2,\sigma_S^2$. If this first
term is not sufficient, localization and dissipation cannot be
assumed to enter additively.

For the particular case of the chain, shown in
Fig.~\ref{fig:JJchain} and described by Eqs.~(\ref{linsys=}),
the dispersion relation of the disorder-free chain is
well-known in the absence of dissipation
($R\to\infty$),\cite{Fazio2001,Rastelli2012,Masluk2012} and can be
straightforwardly generalized to the case of finite~$R$:
\begin{equation}\label{cleandisp=}
\frac{\Omega(k)}{\omega_p}=
\sqrt{\frac{2\ep_k}{2\ep_k+\ell^{-2}}-\frac{\ep_k^2/Q^2}{(2\ep_k+\ell^{-2})^2}}
-i\,\frac{\ep_k/Q}{2\ep_k+\ell^{-2}},
\end{equation}
where we have denoted
\begin{equation}
\ep_k\equiv{2}\sin^2\frac{k}2,\;\;\;
\omega_p\equiv\frac{1}{\sqrt{LC}},\;\;\;
\ell^{-2}\equiv\frac{C^g}C,\;\;\;
Q\equiv\omega_p{RC}.
\end{equation}
Here $\omega_p$ is the plasma frequency of a single junction,
which is a convenient unit of frequency,
$Q$~is the quality factor of a single junction, which is nothing
but the resistance~$R$ in the units of $\sqrt{L/C}$, and
$\ell$~is the screening length, so that $\ell^{-2}$ is the ground
capacitance $C^g$ in the units of~$C$. The real wave vector~$k$
varies from 0 to~$\pi$ for an infinite chain, while for a chain
of $N$~islands assumes $N$ discrete values,
\begin{equation}
\label{k=}
k=0,\frac{\pi}{N},\frac{2\pi}{N},\ldots,\frac{(N-1)\pi}{N},
\end{equation}
the eigenmodes of the system being $V_n\propto\cos{k}(n-1/2)$.
For the low-frequency modes at small $k\ll\min\{1,\ell^{-1}\}$,
\begin{equation}
\frac{\Re\Omega(k)}{\omega_p}\approx{k}\ell,\quad
\frac{\Im\Omega(k)}{\omega_p}\approx-\frac{(k\ell)^2}{2Q},
\end{equation}
so that their damping is weak, $|\Im\Omega(k)|\ll|\Re\Omega(k)|$,
even if the quality factor is not very high.

The inverse absorption length~$\kappa$ for the disorder-free chain
at real~$\omega$ should be found as the solution of the equation
$\Omega(k+i\kappa)=\omega$, which is equivalent to
\begin{equation}\label{cosk=}
1-\cos(k+i\kappa)
=\frac{\varpi^2\ell^{-2}/2}{1-\varpi^2-i\varpi/Q},
\quad\varpi\equiv\frac{\omega}{\omega_p}.
\end{equation}
The resulting expression for $\kappa$ is rather lengthy, so we give
here the approximate expression, valid to the leading order in
$1/Q\ll{1}$:
\begin{equation}\label{kappa=}
\kappa=\frac{1}{Q}\,\frac{\varpi^2\ell^{-1}/2}%
{(1-\varpi^2)\sqrt{1-\varpi^2-\varpi^2\ell^{-2}/4}}+O(Q^{-2}).
\end{equation}
Note that to the leading order in $Q^{-1}$, we have
$\kappa=-\Im\Omega(k)/(d\Re\Omega(k)/dk)$, which corresponds to
the solution of $\Omega(k+i\kappa)=\omega$ perturbatively in the
imaginary parts.

The impedance of a finite disorder-free chain [as defined in
Fig.~\ref{fig:JJchain}(a)] can be represented as a sum over the
eigenmodes (see Sec.~\ref{sec:Eigen} for details):
\begin{subequations}\begin{eqnarray}
\label{Zcleana=}
&&Z(\omega)=\sum_k\left[\frac{iA_k}{\omega-\Omega(k)}
+\frac{iA_k^*}{\omega+\Omega^*(k)}\right],\\
\label{Zcleanb=}
&&A_k=\frac{1}N\,
\frac{4\cos^2(k/2)\sin^2(kN/2)}{C^g+\ep_k[2C+i/\Omega(k)R]}.
\end{eqnarray}\end{subequations}

For a disordered chain, we have calculated the localization
length~$\xi$ in the limit of weak disorder and no dissipation
(see Sec.~\ref{sec:Localization} for details).
As discussed in the beginning of this section, to the leading
order in the disorder strength, one can still label the
eigenstates by their wave vector~$k$, and speak about the
$k$-dependent localization length, which is given by
\begin{equation}\label{xi=}
\frac{1}\xi=\frac{\sigma_S^2+\sigma_g^2}2\,\tan^2\frac{k}2
=\frac{\left(\sigma_S^2+\sigma_g^2\right)\varpi^2\ell^{-2}/8}%
{1-\varpi^2-\varpi^2\ell^{-2}/4}.
\end{equation}
Since $\sigma_R^2,\sigma_g^2\ll{1}$, the inequality $1/\xi\ll{k}$
holds almost everywhere, except for a narrow region of~$k$ around
$k=\pi$. At $\omega\to{0}$ the localization length diverges as
$1/\omega^2$. Such low-frequency behaviour is quite common for
disordered bosonic
problems.\cite{Ziman1982,John1983,Gurarie2003,Bilas2006}
In our case, the divergence is related to the existence of the
delocalized zero mode, {\em i.e.}, an eigenmode with $\omega=0$,
$V_n=\mathrm{const}$, when no currents flow in the system, for
any realization of the disorder. This is the Goldstone mode
related to the global gauge invariance of the system.

The case of short chains, $N\ll\xi,1/\kappa$, is the simplest
to analyze theoretically, and at the same time it is quite
relevant for experiments. Due to the condition $N\ll{1}/\kappa$,
the spacing between eigenmode frequencies is larger than their
broadening, so the discrete modes are well resolved. The
condition $N\ll\xi$ ensures that corrections to the eigenmode
profiles and frequencies are relatively small. The latter,
however, does not mean that the correction to the impedance of
the chain is small. Indeed, the impedance changes significantly
when the eigenmode frequency shift $\delta\omega_k$ due to
disorder is of the order of the broadening $\Im\Omega(k)$, even
if $\delta\omega_k$ is small compared to the frequency
$\Re\Omega(k)$ itself. At the same time, the disorder-induced
corrections to the amplitude $A_k$ and to the broadening produce
just small corrections to the impedance. Thus, we focus on the
random shift~$\delta\omega_k$, calculated perturbatively in
Sec.~\ref{sec:Eigen}. The average
$\langle\delta\omega_k\rangle=0$, and the fluctuations are
given by
\begin{equation}\label{dwk2av=}
\frac{\langle\delta\omega_k^2\rangle}{\omega_p^2}
=\frac{3}8\,\frac{\sigma_S^2+\sigma_g^2}N\,
\frac{2\ep_k\,\ell^{-4}}{(2\ep_k+\ell^{-2})^3}.
\end{equation}

When the chain is long compared to the inverse
Lyapunov exponent, $N\gg(\kappa+1/\xi)^{-1}$, the two ends of
the chain are effectively decoupled. Then the impedance of
the chain~$Z(\omega)$, as defined in Fig.~\ref{fig:JJchain}(a),
equals to the sum of the impedances of two semi-infinite chains,
shown in Fig.~\ref{fig:JJchain}(c). The impedance of a
semi-infinite chain in the absence of disorder, which we denote
by $Z_\infty^{(\mathrm{c})}(\omega)$, is given by
\begin{equation}\begin{split}\label{Zc=}
&Z_\infty^{(\mathrm{c})}(\omega)=\frac{1}{2Y(\omega)}
\left(-1+\sqrt{1-\frac{4Y(\omega)}{i\omega{C}^g}}\right),\\
&Y(\omega)\equiv-\frac{1}{i\omega{L}}+\frac{1}{R}-i\omega{C}.
\end{split}\end{equation}
When disorder is present, the two ends of a sufficiently long
chain feel two different  realizations of disorder, so
$Z(\omega)$ is a sum of two impedances $Z_\infty(\omega)$,
which are statistically independent.
Thus, to characterize the statistics of $Z(\omega)$, it is
sufficient to find the statistics of the impedance
$Z_\infty(\omega)$ of a semi-infinite chain. This impedance is
given by the lower right ({\em i.e.}, $N,N$) element of the inverse
matrix of the system~(\ref{linsys=}).
So far we have made the assumptions of weak dissipation, which
implies $k\gg\kappa$, and of weak disorder, $k\gg{1}/\xi$. Still,
under these assumptions, two possible regimes can be identified:
$\kappa\gg{1}/\xi$ and $\kappa\ll{1}/\xi$. The difference in
statistics of the impedance in the two regimes can be understood
from the following qualitative arguments.

\begin{figure}
\begin{center}
\vspace{0cm}
\includegraphics[width=8cm]{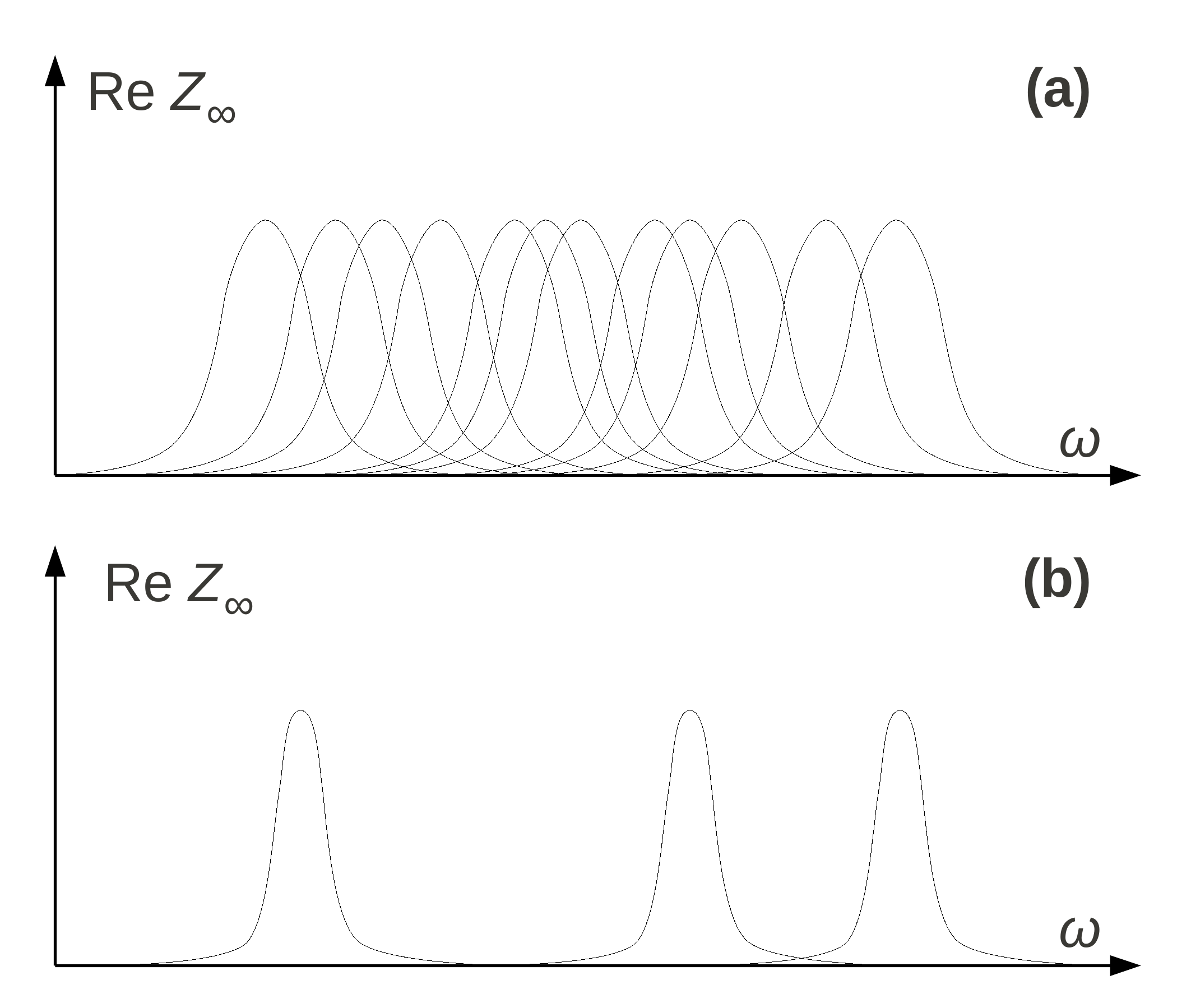}
\vspace{-0cm}
\end{center}
\caption{\label{fig:Lorentzians}
A schematic representation of different terms contributing
to $\Re{Z}_\infty$ (a)~at $\delta_\xi\ll\gamma$ (strongly
overlapping Lorentzians), and (b)~at $\delta_\xi\gg\gamma$
(well-separated Lorentzians).}
\end{figure}

Let us focus on the real part of the impedance, which determines
the absorption. In terms of the complex eigenmode frequencies
which we write as $\Omega_\alpha=\pm\omega_\alpha-i\gamma_\alpha$,
$\alpha=1,\ldots,N-1$, absorption can be represented by a sum of
Lorentzians corresponding to the eigenmodes:
\begin{equation}\label{Lorentzians=}
\Re{Z}_\infty(\omega)=\sum_{\alpha,\pm}
\frac{\gamma_\alpha{A}_\alpha}%
{(\omega\pm\omega_\alpha)^2+\gamma_\alpha^2}
\end{equation}
(the imaginary parts of $A_\alpha$ have been neglected).
Even though the number of terms in the sum can be very large,
only those modes effectively contribute to the sum, which are
located within a distance $\sim\xi$ from the end; for others,
the coefficient $A_\alpha$ is exponentially small. The typical
distance between the Lorentzians corresponds to the
typical frequency spacing $\delta_\xi$ between the modes within
one localization length~$\xi$, while the typical width of each
Lorentzian is~$\gamma$.  Clearly, one can imagine two regimes,
depending on the relation between $\gamma$ and $\delta_\xi$.
When $\delta_\xi\ll\gamma$, the Lorentzians overlap strongly,
so the fluctuations of $Z_\infty(\omega)$ are much smaller
than its average [Fig.~\ref{fig:Lorentzians}(a)]. When
$\delta_\xi\gg\gamma$, the Lorentzians are well separated, and,
depending on the realization of the disorder, the frequency
$\omega$~may either fall near one of the peak
centers~$\omega_\alpha$, which gives a large absorption, or it
may fall between the peaks, and then the absorption will be
small [Fig.~\ref{fig:Lorentzians}(b)]. Thus, in the
case $\delta_\xi\gg\gamma$, the fluctuation will be strong,
so the disorder-averaged impedance is not a very useful concept;
rather, the whole distribution function can be evaluated.
Finally, we recall the relation $\kappa=\gamma/[d\Re\Omega(k)/dk]$,
established earlier, and note that the density of modes within
the localization length can also be evaluated using the
disorder-free dispersion,
$1/\delta_\xi=\xi\,dk/[2\pi\,d\!\Re\Omega(k)]$.
Thus, the parameter controlling the two regimes is precisely
\begin{equation}\label{gammadelta=}
\frac{\gamma}{\delta_\xi}=\frac{\kappa\xi}{2\pi},
\end{equation}
and the mode group velocity $d\Re\Omega(k)/dk$ drops out.

Based on this picture, one can make some estimates. Let us assume
that the main source of fluctuations of $Z_\infty(\omega)$ are the
random positions $\omega_\alpha$. Let us choose an interval of
frequencies, centered at $\omega$ and having the width~$\Delta$,
such that $\delta_\xi,\gamma\ll\Delta\ll\omega$. Let us count
only those modes whose frequencies $\omega_\alpha$ fall inside
this interval; indeed, the modes whose positions $\omega_\alpha$
are too far from~$\omega$ (further than a few times~$\gamma$),
contribute very little to the sum in Eq.~(\ref{Lorentzians=}).
Typically, there are $N_\xi=\Delta/\delta_\xi$ modes inside the
interval. For all such modes, let us set all $\gamma_\alpha$'s
equal to a constant~$\gamma$, and all $A_\alpha=A$ (the smooth
dependence of $A$ and $\gamma$ on~$\omega$ can be neglected due
to the condition $\Delta\ll\omega$). As for the peak positions
$\omega_\alpha$, let us assume them to be uniformly and
independently distributed over the interval
$(\omega-\Delta/2,\omega+\Delta/2)$. Of course, such a Poisson
distribution totally neglects level repulsion, but for our
quantitative estimate is good enough. The average $\Re{Z}_\infty$
is then given by
\begin{equation}\begin{split}
\langle\Re{Z}_\infty\rangle
={}&{}\int\limits_{\omega-\Delta/2}^{\omega+\Delta/2}
\frac{d\omega_1}\Delta\ldots\frac{d\omega_{N_\xi}}\Delta
\sum_{\alpha=1}^{N_\xi}\frac{\gamma{A}}{(\omega-\omega_\alpha)^2+\gamma^2}
=\\
={}&{}\frac{N_\xi}{\Delta}\,\pi{A}
=\frac{\pi{A}}{\delta_\xi}.
\end{split}\end{equation}
Let us now study the probability distribution of the dimensionless
impedance, relative to its average value, which we denote by~$x$:
\begin{equation}\begin{split}\label{Pxdef=}
P(x)={}&{}\int\limits_{\omega-\Delta/2}^{\omega+\Delta/2}
\frac{d\omega_1}\Delta\ldots\frac{d\omega_{N_\xi}}\Delta\times\\
{}&{}\times\delta\!\left(x-\sum_{\alpha=1}^{N_\xi}
\frac{\delta_\xi\gamma/\pi}{(\omega-\omega_\alpha)^2+\gamma^2}
\right).
\end{split}\end{equation}
This probability distribution can be straightforwardly evaluated
in the two limiting cases $\delta_\xi\ll\gamma$ and
$\delta_\xi\gg\gamma$ (see Appendix~\ref{app:distribution} for
details).
\begin{subequations}
For $\delta_\xi\ll\gamma$, we have
\begin{equation}\label{Pxdllg=}
P(x)=\sqrt{\gamma/\delta_\xi}\,e^{-(\pi\gamma/\delta_\xi)(x-1)^2}.
\end{equation}
The narrow Gaussian distribution arises naturally as a consequence
of the central limit theorem, since there are many terms in the
sum~(\ref{Lorentzians=}) which contribute to $\Re{Z}_\infty$.
For $\delta_\xi\gg\gamma$,
\begin{equation}\label{Pxdggg=}
P(x)=\frac{e^{-\gamma/(x\delta_\xi)}}%
{\pi{x}^{3/2}\sqrt{\delta_\xi/(\pi\gamma)-x}},\quad
0<x<\frac{\delta_\xi}{\pi\gamma},
\end{equation}
and $P(x)=0$ outside the indicated interval.
\end{subequations}
The exponential suppression of $P(x)$ at small~$x$ comes from the
fact that an anomalously small~$x$ requires a region of frequencies
of the width $\Delta\omega\gg\delta_\xi$, free of Lorentzians. For
the Poisson distribution of $\omega_\alpha$'s, assumed here, the
probability to have such a region vanishes as
$e^{-\Delta\omega/\delta_\xi}$; if level repulsion is taken into
account, the supression is even stronger. The weak singularity at
large $x=\delta_\xi/(\pi\gamma)$ is a consequence of the assumption
$A_\alpha=A$, $\gamma_\alpha=\gamma$, which implies that all
Lorentzians have the same height. In reality, a spread in the heights
smears the singularity.

In Sec.~\ref{sec:Impedance}, we calculate the distribution function
of $Z_\infty$ in the regime of weak fluctutations ($\kappa\xi\gg{1}$)
using the Fokker-Planck equation:
\begin{equation}\label{PZ=}
P(Z_\infty)\propto\exp\left(-\frac{\kappa\xi}4\,
\frac{|Z_\infty-Z_\infty^{(\mathrm{c})}|^2}%
{|Z_\infty^{(\mathrm{c})}|^2}\right),
\end{equation}
where $\kappa$, $\xi$, and $Z_\infty^{(\mathrm{c})}(\omega)$ are
given by Eqs. (\ref{kappa=}), (\ref{xi=}), and (\ref{Zc=}),
respectively. Keeping in mind the relation~(\ref{gammadelta=}),
we see that the semi-qualitative Eq.~(\ref{PxGauss=}) gives the
correct functional form (Gaussian), underestimating the
fluctuations by a factor of~2. The latter is not surprising, as
in deriving Eq.~(\ref{PxGauss=}) we completely ignored the
fluctuations of the eigenmode amplitudes and widths.
Eq.~(\ref{PZ=}) has also been confirmed by direct numerical
sampling.\cite{Crisan}

In the regime of strong fluctuations, $\kappa\xi\ll{1}$, we were
unable to solve the problem analytically, so we analyzed it
numerically (see Sec.~\ref{sec:Impedance}).
Our numerical results, indeed, indicate that the distribution of
$\Re{Z}_\infty$ has a power-law tail described by
\begin{equation}\label{PZtail=}
P(\Re{Z}_\infty)\approx
C\,\frac{(\kappa\xi)^{0.5}}{\Re{Z}_\infty^{(\mathrm{c})}}
\left(\frac{\Re{Z}_\infty^{(\mathrm{c})}}{\Re{Z}_\infty}\right)^{1.5},
\quad C\sim 0.3-0.5,
\end{equation}
in agreement with Eq.~(\ref{Pxdggg=}). This also agrees with
the distribution of the reflection coefficient, calculated for
the one-dimensional Helmholtz equation with spatially fluctuating
dielectric constant in Ref.~\onlinecite{PradhanKumar}
(see Appendix~\ref{app:reflection} for the relation between the
impedance and the reflection coefficient).

\section{The eigenvalue problem}
\label{sec:Eigen}

To begin with, we note that the formal manipulations performed
in this section, in fact, are quite analogous to those for the
elementary damped harmonic oscillator. The latter is discussed
in Appendix~\ref{app:harmonic} in order to make the present section
more transparent.

Let us write the system~(\ref{linsys=}) in a compact form:
\begin{equation}\label{veclinsys=}
\left(\hat{L}^{-1}-i\omega\hat{R}^{-1}-\omega^2\hat{C}\right)
\vec{V}=-i\omega\vec{I},
\end{equation}
where $\vec{V}$ and $\vec{I}$ are $N$-dimensional column vectors containing the node voltages $V_n$ and currents $I_n$, respectively, and
$\hat{L}^{-1}$, $\hat{R}^{-1}$, and $\hat{C}$ are real, symmetric,
tridiagonal matrices. Besides, they are positive-definite; indeed, for
an arbitrary real vector~$\vec{x}$,
\begin{subequations}\begin{eqnarray}
&&\vec{x}^T\hat{L}^{-1}\vec{x}=
\sum_{n=1}^{N-1}\frac{(x_{n+1}-x_n)^2}{L_{n+1/2}}\geq{0},\\
&&\vec{x}^T\hat{R}^{-1}\vec{x}=
\sum_{n=1}^{N-1}\frac{(x_{n+1}-x_n)^2}{R_{n+1/2}}\geq{0},\\
&&\vec{x}^T\hat{C}\vec{x}=\sum_{n=1}^{N-1}C_{n+1/2}(x_{n+1}-x_n)^2
+\sum_{n=1}^NC^g_nx_n^2>0.\nonumber\\ &&
\end{eqnarray}\end{subequations}
$\hat{L}^{-1}$ and $\hat{R}^{-1}$ have exactly one zero eigenvalue with
the eigenvector $\vec{x}_0=(1\:1\ldots{1})^T/\sqrt{N}$. At the same
time, $\vec{x}_0^T\hat{C}\vec{x}_0=\overline{C^g}$, the spatial average
of the ground capacitance.

Thanks to these properties, one can define the matrix square roots
$\hat{L}^{-1/2},\hat{C}^{1/2}$, as well as the inverse $\hat{C}^{-1/2}$,
which are also real, symmetric, and positive-definite matrices. Then
the system~(\ref{veclinsys=}), quadratic in~$\omega$ can be identically
rewritten as
\begin{widetext}\begin{equation}\label{C12vw=}
\omega\mathcal{V}-\left(\begin{array}{cc}
-i\hat{C}^{-1/2}\hat{R}^{-1}\hat{C}^{-1/2}
& \hat{C}^{-1/2}\hat{L}^{-1/2} \\
\hat{L}^{-1/2}\hat{C}^{-1/2} & 0 \end{array}\right)\mathcal{V}=
\left(\begin{array}{c} i\hat{C}^{-1/2}\vec{I} \\ 0 \end{array}\right),
\quad
\mathcal{V}\equiv\left(\begin{array}{c} \hat{C}^{1/2}\vec{V} \\
\vec{w} \end{array}\right)
\end{equation}\end{widetext}
where $\vec{w}$ is an auxiliary $N$-dimensional column vector.
It should be noted that while $\hat{L}^{-1}$, $\hat{R}^{-1}$, and
$\hat{C}$ are tridiagonal and thus are relatively easy to deal with
numerically, their square roots are non-local, so Eq.~(\ref{C12vw=})
is only convenient for formal manipulations. The main advantage of
Eq.~(\ref{C12vw=}) is that its solutions can be expressed in terms of
eigenvalues and egenvectors of the $2N\times{2}N$ matrix
\begin{equation}\label{matrixA=}
\hat{\mathcal{A}}=\left(\begin{array}{cc}
-i\hat{C}^{-1/2}\hat{R}^{-1}\hat{C}^{-1/2}
& \hat{C}^{-1/2}\hat{L}^{-1/2} \\
\hat{L}^{-1/2}\hat{C}^{-1/2} & 0 \end{array}\right)
\end{equation}
The matrix $\hat{\mathcal{A}}$ is non-Hermitian, so it may have less
then $2N$ eigenvectors if its eigenvalues are degenerate.
In a disordered system all degeneracies can be assumed to be lifted,
except those which are protected by a symmetry and do not depend on
the disorder realization.
The only special value of $\omega$ is $\omega=0$, due to the gauge
invariance, as discussed in Sec.~\ref{sec:Discussion}. For any
disorder realization, it is an eigenvalue with two eigenvectors:
\begin{equation}
\mathcal{V}_{0\pm}=\frac{1}{\sqrt{2\overline{C^g}}}
\left(\begin{array}{c} \hat{C}^{1/2}\vec{x}_0 \\
\pm{i}\sqrt{\overline{C^g}}\vec{x}_0 \end{array}\right).
\end{equation}
Thus, with probability~1, the matrix $\hat{\mathcal{A}}$ has exactly
$2N$ eigenvectors, which form a complete set. Since it is symmetric,
$\hat{\mathcal{A}}^T=\hat{\mathcal{A}}$, different eigenvectors
$\alpha$ and $\beta$ are orthogonal as
\begin{equation}
\mathcal{V}^T_\alpha\mathcal{V}_\beta=
\vec{V}^T_\alpha\hat{C}\vec{V}_\beta
+\vec{w}^T_\alpha\vec{w}_\beta=\delta_{\alpha\beta}.
\end{equation}
The matrix $\hat{\mathcal{A}}$ satisfies the property
$\sigma_z\hat{\mathcal{A}}^*\sigma_z=-\hat{\mathcal{A}}$, where
$\sigma_z$ is the Pauli matrix acting in the $2\times{2}$ space of
$N\times{N}$ blocks. Thus, if $\omega$ is an eigenvalue of
$\mathcal{A}$ with the corresponding eigenvector
$(\hat{C}^{1/2}\vec{V}\;\vec{w})^T$, then $-\omega^*$ is also an
eigenvalue, and the corresponding eigenvector is
$(\hat{C}^{1/2}\vec{V}^*\;-\vec{w}^*)^T$.
Since the absorption is assumed to be weak, we neglect the
possibility to have purely imaginary eigenvalues, and count by
$\alpha=1,\ldots,N$ the eigenvalues with $\Re\omega_\alpha>0$.
Then the unit $2N\times{2}N$ matrix can be represented as
\begin{equation}\begin{split}
\hat{\openone}={}&{}\left(\begin{array}{cc}
\hat{C}^{1/2}\vec{x}_0\vec{x}_0^T\hat{C}^{1/2}/\overline{C^g} &
0 \\ 0 & \vec{x}_0\vec{x}_0^T\end{array}\right){}+{}\\
{}&{}+\sum_{\alpha=1}^{N-1}
\left(\begin{array}{cc}
\hat{C}^{1/2}\vec{V}_\alpha\vec{V}_\alpha^T\hat{C}^{1/2} &
\hat{C}^{1/2}\vec{V}_\alpha\vec{w}_\alpha^T \\
\vec{w}_\alpha\vec{V}_\alpha^T\hat{C}^{1/2} &
\vec{w}_\alpha\vec{w}_\alpha^T\end{array}\right){}+{}\\
&{}+{}\sum_{\alpha=1}^{N-1}
\left(\begin{array}{cc}
\hat{C}^{1/2}\vec{V}_\alpha^*\vec{V}_\alpha^\dagger\hat{C}^{1/2} &
-\hat{C}^{1/2}\vec{V}_\alpha^*\vec{w}_\alpha^\dagger \\
-\vec{w}_\alpha^*\vec{V}_\alpha^\dagger\hat{C}^{1/2} &
\vec{w}_\alpha^*\vec{w}_\alpha^\dagger \end{array}\right),
\end{split}\end{equation}
and the resolvent of $\hat{\mathcal{A}}$ as
\begin{equation}\begin{split}
&(\omega-\hat{\mathcal{A}})^{-1}
=\frac{1}\omega\left(\begin{array}{cc}
\hat{C}^{1/2}\vec{x}_0\vec{x}_0^T\hat{C}^{1/2}/\overline{C^g} &
0 \\ 0 & \vec{x}_0\vec{x}_0^T\end{array}\right){}+{}\\
&{}+{}\sum_{\alpha=1}^{N-1}\frac{1}{\omega-\omega_\alpha}
\left(\begin{array}{cc}
\hat{C}^{1/2}\vec{V}_\alpha\vec{V}_\alpha^T\hat{C}^{1/2} &
\hat{C}^{1/2}\vec{V}_\alpha\vec{w}_\alpha^T \\
\vec{w}_\alpha\vec{V}_\alpha^T\hat{C}^{1/2} &
\vec{w}_\alpha\vec{w}_\alpha^T\end{array}\right){}+{}\\
&{}+{}\sum_{\alpha=1}^{N-1}\frac{1}{\omega+\omega_\alpha^*}
\left(\begin{array}{cc}
\hat{C}^{1/2}\vec{V}_\alpha^*\vec{V}_\alpha^\dagger\hat{C}^{1/2} &
-\hat{C}^{1/2}\vec{V}_\alpha^*\vec{w}_\alpha^\dagger \\
-\vec{w}_\alpha^*\vec{V}_\alpha^\dagger\hat{C}^{1/2} &
\vec{w}_\alpha^*\vec{w}_\alpha^\dagger \end{array}\right).
\end{split}\end{equation}
Eliminating the auxiliary vector~$\vec{w}$, we obtain
\begin{equation}
\vec{V}=\left(\frac{\vec{x}_0\vec{x}_0^T}{\omega\overline{C^g}}
+\sum_{\alpha=1}^{N-1}
\frac{\vec{V}_\alpha\vec{V}_\alpha^T}{\omega-\omega_\alpha}
+\sum_{\alpha=1}^{N-1}
\frac{\vec{V}_\alpha^*\vec{V}_\alpha^\dagger}{\omega+\omega_\alpha^*}
\right)i\vec{I},
\end{equation}
where the eigenvectors $\vec{V}_\alpha$ should be normalized as
\begin{equation}\label{Vnorm=}
\vec{V}_\alpha^T\left(\hat{C}+\omega_\alpha^{-2}\hat{L}^{-1}\right)
\vec{V}_\alpha=1,\quad\alpha=1,\ldots,N-1.
\end{equation}
The impedance of a finite chain as defined in
Fig.~\ref{fig:JJchain}(a), is now given by
\begin{equation}\label{ZV=}
Z(\omega)=i\sum_{\alpha=1}^{N-1}\left[
\frac{(V_{\alpha{1}}-V_{\alpha{N}})^2}{\omega-\omega_\alpha}
+\frac{(V_{\alpha{1}}^*-V_{\alpha{N}}^*)^2}{\omega+\omega_\alpha^*}
\right].
\end{equation}
In a disorder-free chain the eigenmodes are plane waves ({\em c.f.} Eq.~\ref{k=}):
\begin{subequations}\begin{equation}\label{eigenmode=}
V_{kn}=a_k\cos{k}(n-1/2),\quad
k=0,\frac{\pi}{N},\quad\ldots,(N-1)\,\frac{\pi}{N},
\end{equation}
where the amplitudes $a_k$ are determined by the normalization
condition~(\ref{Vnorm=}):
\begin{equation}\begin{split}
\frac{1}{a_{k\neq{0}}^2}{}&{}=
\frac{N}2\,C^g+N\ep_k\left(C+\frac{1}{\omega_k^2L}\right)=\\
{}&{}=NC^g+N\ep_k\left(2C+\frac{i}{\omega_kR}\right).
\end{split}\end{equation}\end{subequations}
Substitution of these expressions into Eq.~(\ref{ZV=}) gives
Eqs.~(\ref{Zcleana=}),~(\ref{Zcleanb=}). Here and below the
following relations prove useful:
\begin{equation}\begin{split}\label{sumcos=}
&\sum_{n=1}^N\cos^2k(n-1/2)=\frac{N}2,\\
&\sum_{n=1}^N\cos^4k(n-1/2)=\frac{3N}8,\\
&\sum_{n=1}^{N-1}[\cos{k}(n+1/2)-\cos{k}(n-1/2)]^2=
N\ep_k,\\
&\sum_{n=1}^{N-1}[\cos{k}(n+1/2)-\cos{k}(n-1/2)]^4=
\frac{3N}{2}\,\ep_k^2.
\end{split}\end{equation}

Let us determine the shift of an eigenvalue $\omega_\alpha$ due
to a small perturbation. If $\delta\hat{\mathcal{A}}$ is a
perturbation of $\hat{\mathcal{A}}$, the shift is given by
\begin{equation}
\delta\omega_\alpha=
\mathcal{V}_\alpha^T\,\delta\hat{\mathcal{A}}\,\mathcal{V}_\alpha.
\end{equation}
Assuming that the perturbation is due to a fluctuation in the
capacitances and inductances and neglecting this fluctuation
when it is multiplied by a small quantity $1/R$, we can write
\begin{equation}
\delta\hat{\mathcal{A}}=\left(\begin{array}{cc}
0 & \delta(\hat{C}^{-1/2}\hat{L}^{-1/2}) \\
\delta(\hat{L}^{-1/2}\hat{C}^{-1/2}) & 0
\end{array}\right),
\end{equation}
which gives
\begin{equation}
\delta\omega_\alpha=\frac{1}{\omega_\alpha}\,\vec{V}_\alpha^T
\left(\delta\hat{L}^{-1}-\omega_\alpha^2\delta\hat{C}\right)
\vec{V}_\alpha.
\end{equation}
For the fluctuations given by Eq.~(\ref{disorder=}), this becomes
\begin{equation}\begin{split}
\delta\omega_\alpha={}&{}
\left(\frac{1}{\omega_\alpha{L}}-\omega_\alpha{C}\right)
\sum_{n=1}^{N-1}\zeta_n(V_{\alpha,n+1}-V_{\alpha,n})^2{}-{}\\
{}&{}-\omega_\alpha{C}^g\sum_{n=1}^{N-1}\eta_nV_{\alpha,n}^2.
\end{split}\end{equation}
For the eigenmodes (\ref{eigenmode=}) the fluctuation
$\langle\delta\omega_k^2\rangle$ can be evaluated using
relations (\ref{sumcos=}), which gives Eq.~(\ref{dwk2av=}).

\section{Localization length of the normal modes}
\label{sec:Localization}

To determine the localization length $\xi$ for the eigenmodes of
the system~(\ref{linsys=}) in the absence of dissipation ($R\to\infty$),
we rewrite it identically in the form
\begin{equation}
\underline{v_{n+1}}=\underline{\underline{m_n}}\,\underline{v_n},
\end{equation}
where $\underline{v_n}$ is two-component column
\begin{equation}
\underline{v_n}=\sqrt{Y_{n-1/2}}\left(\begin{array}{c} V_n\\ V_{n-1}\end{array}\right),
\end{equation}
and $\underline{\underline{m_n}}$ is the transfer matrix
\begin{equation}\begin{split}
&\underline{\underline{m_n}}=
\left(\begin{array}{cc} y_n+1/y_n
-\frac{i\omega{C}^g_n}{\sqrt{Y_{n-1/2}Y_{n+1/2}}}
& -1/y_n \\ y_n & 0 \end{array}\right),\\
&y_n\equiv\sqrt{\frac{Y_{n+1/2}}{Y_{n-1/2}}}.
\end{split}\end{equation}
The localization length is calculated from the Lyapunov exponent
of the product
$\underline{\underline{m_n}}\ldots\underline{\underline{m_1}}$
at $n\to\infty$ (see, e.~g., Ref.~\onlinecite{LGP}).

We substitute the expressions from Eq.~(\ref{disorder=}),
and expand to second order and omit products of uncorrelated
fluctuations:
\begin{equation}
\underline{\underline{m_n}}=
\left(\begin{array}{cc} 2(1-\ep)(1+\Xi_n^2/2)+\ep\Upsilon_n
& -1+\Xi_n-\Xi_n^2/2 \\ 1+\Xi_n+\Xi_n^2/2 & 0 \end{array}\right),
\end{equation}
where we have denoted for brevity
\begin{subequations}\begin{eqnarray}
&&\Xi_n=\frac{\zeta_n-\zeta_{n-1}}2-\frac{\zeta_n^2-\zeta_{n-1}^2}4,\\
&&\Upsilon_n=\zeta_n+\zeta_{n-1}-2\eta_n
-\frac{\zeta_n^2+\zeta_{n-1}^2}2,
\end{eqnarray}\end{subequations}
and
\begin{equation}
\ep=\frac{\omega^2LC^g/2}{1-\omega^2LC}
\end{equation}
has the same meaning as in Sec.~\ref{sec:Discussion}: for such
real values of~$\omega$ that $0\leq\ep\leq{2}$, the
solution of the equation $\ep=1-\cos{k}$ determines the
dispersion relation of the disorder-free chain. This is precisely
the range of frequencies we are interested in.

Let us switch to the basis in which the the evolution of $\underline{v_n}$
for the disorder-free chain is trivial:
\begin{equation}
\underline{v_n}=\underline{\underline{K_n}}\,\underline{\tilde{v}_n},\quad
\underline{\underline{K_n}}
\equiv\left(\begin{array}{cc} e^{ikn} & e^{-ikn} \\
e^{ik(n-1)} & e^{-ik(n-1)} \end{array}\right).
\end{equation}
The rotated transfer matrix $\underline{\underline{\tilde{m}_n}}
=\underline{\underline{K_{n+1}^{-1}}}\,\underline{\underline{m_n}}\,
\underline{\underline{K_n}}$ is given by
\begin{equation}\begin{split}
\underline{\underline{\tilde{m}_n}}={}&{}1+\frac{\Xi_n^2}2
+\Xi_n\left(\begin{array}{cc} 0 & e^{-2ikn} \\
e^{2ikn} & 0 \end{array}\right)+\\
{}&{}+\frac{\Upsilon_n}{2i}\tan\frac{k}2
\left(\begin{array}{cc} 1 & e^{-2ikn} \\ -e^{2ikn} & -1 \end{array}\right).
\end{split}\end{equation}
In the disorder-free case the two components of the vector
$\underline{\tilde{v}_n}$ represent the amplitudes of the
right- and left-travelling waves. The product of the transfer
matrices
$\underline{\underline{\tilde{m}_n}}\ldots\underline{\underline{\tilde{m}_1}}$
of a disordered segment of length~$n$ determines the amplitude
reflection and transmission coefficients of this segment.
The Hermitian matrix
\begin{equation}
\underline{\underline{M_n}}=
\underline{\underline{\tilde{m}_n}}\ldots
\underline{\underline{\tilde{m}_1}}\,
\underline{\underline{\tilde{m}_1^\dagger}}\ldots
\underline{\underline{\tilde{m}_n^\dagger}}
\end{equation}
determines the intensity transmission coefficient.
Namely, by noting that $\underline{\underline{M_n}}$ satisfies
the constraints
\begin{equation}
\underline{\underline{M_n}}=\underline{\underline{M_n^\dagger}},\quad
\det\underline{\underline{M_n}}=1,\quad
\underline{\underline{M_n}}=
\sigma_x\underline{\underline{M_n^*}}\sigma_x,
\end{equation}
where $\sigma_x$ is the first Pauli matrix, and the last two constraints
follow from the same properties obeyed by $\underline{\underline{\tilde{m}_n}}$
one can parametrize the matrix $\underline{\underline{M_n}}$ as
\begin{equation}
\underline{\underline{M_n}}=
\left(\begin{array}{cc} \cosh\mu_n & e^{i\phi_n}\sinh\mu_n \\
e^{i\phi_n}\sinh\mu_n & \cosh\mu_n \end{array}\right),\quad\mu_n\geq{0}.
\mu_n\geq{0}.
\end{equation}
The eigenvalues of $\underline{\underline{M_n}}$ are $e^{\pm\mu_n}$,
and the intensity transmission coefficient of the disordered segment
of $n$ sites is $1/\cosh^2(\mu_n/2)$ (Ref.~\onlinecite{MelloKumar}).
At $n\to\infty$, it should decrease exponentially as $e^{-2n/\xi}$,
where $\xi$ is the localization length of the envelope wave function.
Thus, $\xi$ can be determined from the relation
\begin{equation}
\frac{1}\xi=\frac{1}{2}\lim_{n\to\infty}\frac{\mu_n}{n}.
\end{equation}
Note that statistical averaging is not needed here: $\mu_n$ is a
self-averaging quantity\cite{LGP}, as naturally follows from the fact
that it is a logarithm of the product of many independent factors.

The change of the matrix~$\underline{\underline{M_n}}$ upon one
iteration is given by
\begin{widetext}\begin{equation}\begin{split}
\underline{\underline{M_{n+1}}}{}={}&{}\underline{\underline{M_n}}
+2\Xi_n\left(\begin{array}{cc} \cos\varphi_n\sinh\mu_n &
e^{-2ikn}\cosh\mu_n \\ e^{2ikn}\cosh\mu_n & \cos\varphi_n\sinh\mu_n
\end{array}\right)+\\
{}&{}+{}\frac{\ep\Upsilon_n}{\sin{k}}\left(\begin{array}{cc}
-\sin\varphi_n\sinh\mu_n &
-ie^{-2ikn}[\cosh\mu_n+e^{i\varphi_n}\sinh\mu_n] \\
ie^{2ikn}[\cosh\mu_n+e^{-i\varphi_n}\sinh\mu_n] &
-\sin\varphi_n\sinh\mu_n \end{array}\right)+\\
{}&{}+{}2\Xi_n^2\left(\begin{array}{cc}
\cosh\mu_n & e^{-2ikn}\cos\varphi_n\sinh\mu_n \\
e^{2ikn}\cos\varphi_n\sinh\mu_n & \cosh\mu_n\end{array}\right)
+\\ &{}+
\frac{\ep\Upsilon_n\Xi_n}{\sin{k}}\left(\begin{array}{cc}
\sin\varphi_n\sinh\mu_n &
-ie^{-2ikn}[\cosh\mu_n+e^{-i\varphi_n}\sinh\mu_n] \\
ie^{2ikn}[\cosh\mu_n+e^{i\varphi_n}\sinh\mu_n] &
\sin\varphi_n\sinh\mu_n \end{array}\right)
+\\ &{}+
\frac{\ep^2\Upsilon_n^2[\cosh\mu_n+\cos\varphi_n\sinh\mu_n]}{2\sin^2k}
\left(\begin{array}{cc} 1 & -e^{-2ikn} \\ -e^{2ikn} & 1 \end{array}\right).
\end{split}\end{equation}\end{widetext}
where we denoted $\varphi_n=2kn+\phi_n$.
Let us also denote $\ep\Upsilon_n/\sin{k}=2\tilde\Upsilon_n$.
Then
\begin{subequations}\begin{eqnarray}\label{mun+1=}
&&\mu_{n+1}=\mu_n+2\Xi_n\cos\varphi_n
-2\tilde\Upsilon_n\sin\varphi_n+\nonumber\\
&&\qquad\quad{}+2(\Xi_n\sin\varphi_n+\tilde\Upsilon_n\cos\varphi_n)^2
\coth\mu_n+\nonumber\\
&&\qquad\quad{}+2(\Xi_n\sin\varphi_n
+\tilde\Upsilon_n\cos\varphi_n)\tilde\Upsilon_n,\\
\label{phin+1=}
&&\phi_{n+1}=\phi_n-2(\Xi_n\sin\varphi_n
+\tilde\Upsilon_n\cos\varphi_n)\coth\mu_n-2\tilde\Upsilon_n+\nonumber\\
&&\qquad\quad{}+O(\Xi_n^2,\tilde\Upsilon_n^2,\Xi_n\tilde\Upsilon_n).
\end{eqnarray}\end{subequations}
In fact,
to determine the average growth rate of $\mu_n$, we do not need
to know the dynamics of $\phi_n$, because the phase $\varphi_n$
entering Eq.~(\ref{mun+1=}) contains the term $2kn$ that varies
more rapidly than~$\phi_n$. Indeed, the evolution of $\phi_n$
is governed by the disorder, and thus occurs on the typical length
scale $\xi$, while for $2kn$ this scale is $\sim{1}/k$. As discussed
in Sec.~\ref{sec:Discussion}, we are interested in the weak-disorder
limit, $k\xi\gg{1}$ (otherwise the whole approach of this section
is not valid). Thus, summation of many small increments
$(\mu_{n+1}-\mu_n)+(\mu_{n+2}-\mu_{n+1})+\ldots+(\mu_{n+l}-\mu_{n+l-1})$
for $l\gg{1}/k$ is equivalent to averaging over $\varphi$
(see Ref.~\onlinecite{commensurate}, though).

It is important to note that $\Xi_n,\Upsilon_n$ are not independent of
$\Xi_{n-1},\Upsilon_{n-1}$. Because of this, it is not sufficient
just to average the right-hand side of Eq.~(\ref{mun+1=}) over the
disorder and over the phase~$\varphi_n$. Indeed, expressing $\phi_n$
in terms of $\phi_{n-1}$ with the help of Eq.~(\ref{phin+1=}), we
obtain (up to the second order in $\Xi,\Upsilon$):
\[\begin{split}
&2\Xi_n\cos\varphi_n
-2\tilde\Upsilon_n\sin\varphi_n=\\
&=2\Xi_n\cos(\varphi_{n-1}+2k)
-2\tilde\Upsilon_n\sin(\varphi_{n-1}+2k)+{}\\
&\;\;+2\tilde\Upsilon_{n-1}[2\Xi_n\sin(\varphi_{n-1}+2k)
+2\tilde\Upsilon_n\cos(\varphi_{n-1}+2k)]+{}\\
&\;\;+[2\Xi_n\sin(\varphi_{n-1}+2k)
+2\tilde\Upsilon_n\cos(\varphi_{n-1}+2k)]\times{}\\
&\quad\times[2\Xi_{n-1}\sin\varphi_{n-1}
+2\tilde\Upsilon_{n-1}\cos\varphi_{n-1}].
\end{split}\]
The average of the last term over the phase $\varphi_{n-1}$ is
not zero and should be taken into account.

Finally, we assume $\mu_n\gg{1}$ and set $\coth\mu_n\to{1}$.
Then, the average increment of $\mu_n$ in one step is given by
\begin{equation}\begin{split}
\frac{2}\xi{}={}&{}\langle\Xi_n^2\rangle+\langle\tilde\Upsilon_n^2\rangle
+2\langle\Xi_n\Xi_{n-1}+\tilde\Upsilon_n\tilde\Upsilon_{n-1}\rangle\cos{2}k
{}+{}\\&{}+{}2\langle\Xi_n\tilde\Upsilon_{n-1}
-\tilde\Upsilon_n\Xi_{n-1}\rangle\sin{2}k,
\end{split}\end{equation}
which yields Eq.~(\ref{xi=}).

\section{Impedance fluctuations}
\label{sec:Impedance}

In order to calculate the probability distribution of the impedance
of a semi-infinite chain, defined in Fig.~\ref{fig:JJchain}(c), let
us consider $Z_N(\omega)$, the impedance of a chain with $N-1$
junctions, but defined according to Fig.~\ref{fig:JJchain}(c),
instead of Fig.~\ref{fig:JJchain}(a). Upon adding one junction to
the chain, its impedance is transformed as
\begin{equation}\label{recursive=}
Z_{N+1}=\frac{1}{-i\omega{C}_{N+1}^g+(Z_N+1/Y_{N+1/2})^{-1}},
\end{equation}
where
\begin{equation}\label{Yw=}
Y_{N+1/2}(\omega)=-\frac{1}{i\omega{L}_{N+1/2}}+
\frac{1}{R_{N+1/2}}-i\omega{C}_{N+1/2}
\end{equation}
is the admittance of the added junction.

In the disorder-free chain, the recursive relation~(\ref{recursive=})
has a stationary point $Z_\infty^{(\mathrm{c})}(\omega)$, given by
Eq.~(\ref{Zc=}). In the vicinity of this stationary point, the
recursive relation can be linearized,
\begin{equation}\label{linrecursive=}
Z_{N+1}-Z_\infty^{(\mathrm{c})}\approx
\tau\left[Z_N-Z_\infty^{(\mathrm{c})}\right],
\end{equation}
with the eigenvalue~$\tau$ given by
\begin{equation}
\tau=\left(
\frac{YZ_\infty^{(\mathrm{c})}}{YZ_\infty^{(\mathrm{c})}+1}\right)^2
=e^{2ik-2\kappa},
\end{equation}
where $k,\kappa$ are defined as solutions of the equation
$1-\cos(k+i\kappa)=i\omega{C}^g/[2Y(\omega)]$, which is identical
to Eq.~(\ref{cosk=}).
Thus, the effect of dissipation, $\kappa>0$, is to squeeze the
points $Z_N$ towards the stationary point in the complex plane.

In the presence of a weak disorder, the recursive relation becomes
random. The effect of the randomness is to make $Z_N$ perform a
random walk in the complex plane, thereby taking them away from the
stationary point. Thus, disorder and dissipation are competing.
At $N\to\infty$ they balance each other, and the probability
distribution of $Z_N$ reaches a stationary limit. If dissipation is
strong enough compared to disorder, the stationary distribution is
concentrated near the stationary point, where the linearized
recursive relation~(\ref{linrecursive=}) is valid. Below we will
calculate the stationary distribution for this case analytically,
and show that it is Gaussian. This case corresponds precisely to
the limit $\kappa\xi\gg{1}$ discussed in Sec.~\ref{sec:Discussion}.

Let us take into account fluctuations of $C^g_{N+1}$ and $Y_{N+1/2}$,
determined by Eq.~(\ref{disorder=}). They produce an additional
stochastic term in Eq.~(\ref{linrecursive=}). To the second order in
$\eta_{N+1},\zeta_N$, the linearized recursive relation becomes
\begin{equation}\begin{split}
&Y\left[Z_{N+1}-Z_\infty^{(\mathrm{c})}\right]=
e^{2ik-2\kappa}Y\left[Z_N-Z_\infty^{(\mathrm{c})}\right]-{}\\
&\qquad{}-e^{ik}\eta_{N+1}-e^{2ik}\zeta_N
+\frac{ie^{5ik/2}\zeta_N^2}{2\sin(k/2)}-{}\\
&\qquad{}-2ie^{3ik/2}\sin\frac{k}2\left(\eta_{N+1}^2
+e^{ik}\eta_{N+1}\zeta_N+e^{2ik}\zeta_N^2\right),
\end{split}\end{equation}
where $Y$ is the same as for the clean chain [{\em i.e.}, defined
as in Eq.~(\ref{Yw=}), but in terms of the non-fluctuating quantities
$L,C,R$].
We neglected $\kappa$ in the coefficients at $\eta_{N+1},\zeta_N$,
as $\kappa\ll{1}$ gives just a small correction to the diffusion
produced by the stochastic terms.

Let us write
\begin{equation}\label{YZlin=}
Y\left[Z_N-Z_\infty^{(\mathrm{c})}\right]=i\sqrt{\rho_N}\,e^{2ikN+i\phi_N}.
\end{equation}
We calculate the increment $\rho_{N+1}-\rho_N$ to the first order in
$\kappa$ and to the second order in $\eta_{N+1},\zeta_N$:
\begin{widetext}\begin{subequations}\begin{eqnarray}\label{rN+1=}
\rho_{N+1}-\rho_N&=&-4\kappa{r}_N
+\sqrt{\rho_N}\,\eta_{N+1}\,2\sin(2kN+\phi_N+k)
+\sqrt{\rho_N}\,\zeta_N\,2\sin(2kN+\phi_N)-\nonumber\\
&&{}-\sqrt{\rho_N}\,\eta_{N+1}^2\,
4\sin\frac{k}2\cos\left(2kN+\phi_N+\frac{k}2\right)
-\sqrt{\rho_N}\,\eta_{N+1}\zeta_N\,
4\sin\frac{k}2\cos\left(2kN+\phi_N-\frac{k}2\right)-\nonumber\\
&&{}-\sqrt{\rho_N}\,\zeta_N^2\,
4\sin\frac{k}2\cos\left(2kN+\phi_N-\frac{3k}2\right)
+\sqrt{\rho_N}\,\zeta_N^2\,\frac{\cos(2kN+\phi_N-k/2)}{\sin(k/2)}+\nonumber\\
&&{}+2\eta_{N+1}\zeta_N\cos{k}+\eta_{N+1}^2+\zeta_N^2,\\
\phi_{N+1}-\phi_N&=&O(\eta_{N+1},\zeta_N).
\end{eqnarray}\end{subequations}\end{widetext}
To determine the stationary distribution of $Z_N$, we do not need
to know the dynamics of $\phi_N$. Indeed, the main contribution to
the dynamics of the phase of $Z_N-Z_\infty^{(\mathrm{c})}$ comes
from the factor $e^{2ikN}$ that we have explicitly separated in
Eq.~(\ref{YZlin=}). The trigonometric factors $\cos(2kN+\ldots)$,
$\sin(2kN+\ldots)$ in Eq.~(\ref{rN+1=}) average to zero after
$N\sim{1}/k$ steps\cite{commensurate} (note that even for $k\ll{1}$,
when this averaging length is large, the length of the chain,
$N\gg{1}/\kappa$ is still larger since $\kappa\ll{k}$).
Hence, the drift of $\rho_N$, determined by the
average of the right-hand side of Eq.~(\ref{rN+1=}), is
contributed to only by the first and the two last terms of
Eq.~(\ref{rN+1=}). The diffusion is determined by the average
square of the right-hand side of Eq.~(\ref{rN+1=}), so it is
contributed to by the second and the third terms, which are linear
in $\eta_{N+1}$ and $\zeta_N$ (all quadratic terms give a
higher-order contribution). Following the standard
procedure\cite{VanKampen}, one arrives at the Fokker-Plank
equation for the probability distribution $P_N(\rho)$:
\begin{equation}
\frac{\partial{P}}{\partial{N}}=\frac\partial{\partial\rho}\,\rho
\left[4\kappa{P}+(\sigma_S^2+\sigma_g^2)\,
\frac{\partial{P}}{\partial\rho}\right].
\end{equation}
From its solution,
\begin{equation}\label{PNr=}
P_N(\rho)\propto\exp\left[-\frac{4\kappa\rho/(\sigma_S^2+\sigma_g^2)}%
{1-e^{-4\kappa(N-N_0)}}\right],
\end{equation}
which in the stationary limit ($N\to\infty$) reduces to Eq.~(\ref{PZ=}),
we also extract the typical length, $1/(4\kappa)$, at which this
stationary limit is reached.

In the case $\kappa\xi\ll{1}$, we were unable to obtain an analytical
solution. To treat the problem numerically, and in particular,
to analyze the power-law tail of the distribution of $\Re{Z}_\infty$,
discussed in Sec.~\ref{sec:Discussion}, it is convenient to introduce
the logarithmic variable
\begin{equation}
\lambda=\ln
\frac{\Re{Z}_\infty(\omega)}{\Re{Z}_\infty^{(\mathrm{c})}(\omega)},
\end{equation}
where $\Re{Z}_\infty^{(\mathrm{c})}(\omega)$ is the impedance of
a semi-infinite disorder-free chain, introduced in Eq.~(\ref{Zc=}),
and used here as a convenient unit of measure. If the ratio
$\Re{Z}_\infty/\Re{Z}_\infty^{(\mathrm{c})}$ has a power-law
distribution,
\begin{equation}\label{powerlaw=}
P(\Re{Z}_\infty/\Re{Z}_\infty^{(\mathrm{c})})=
A\left(\frac{\Re{Z}_\infty^{(\mathrm{c})}}{\Re{Z}_\infty}\right)^\alpha,
\end{equation}
in some range of $\Re{Z}_\infty$, the corresponding distribution
of~$\lambda$ is exponential,
$P(\lambda)=Ae^{-(\alpha-1)\lambda}$. Thus, in the following
we study numerically $\ln{P}(\lambda)$, and extract the
exponent~$\alpha$ and the prefactor~$A$ from the slope and the offset
of the dependence $\ln{P}(\lambda)$ versus~$\lambda$.
Each distribution is obtained from about $10^6-10^7$ realizations of the
chain, and the convergence of the limit $N\to\infty$ is reached at
$N\gtrsim(5-6)\,\xi$.

\begin{figure}
\begin{center}
\vspace{0cm}
\includegraphics[width=8cm]{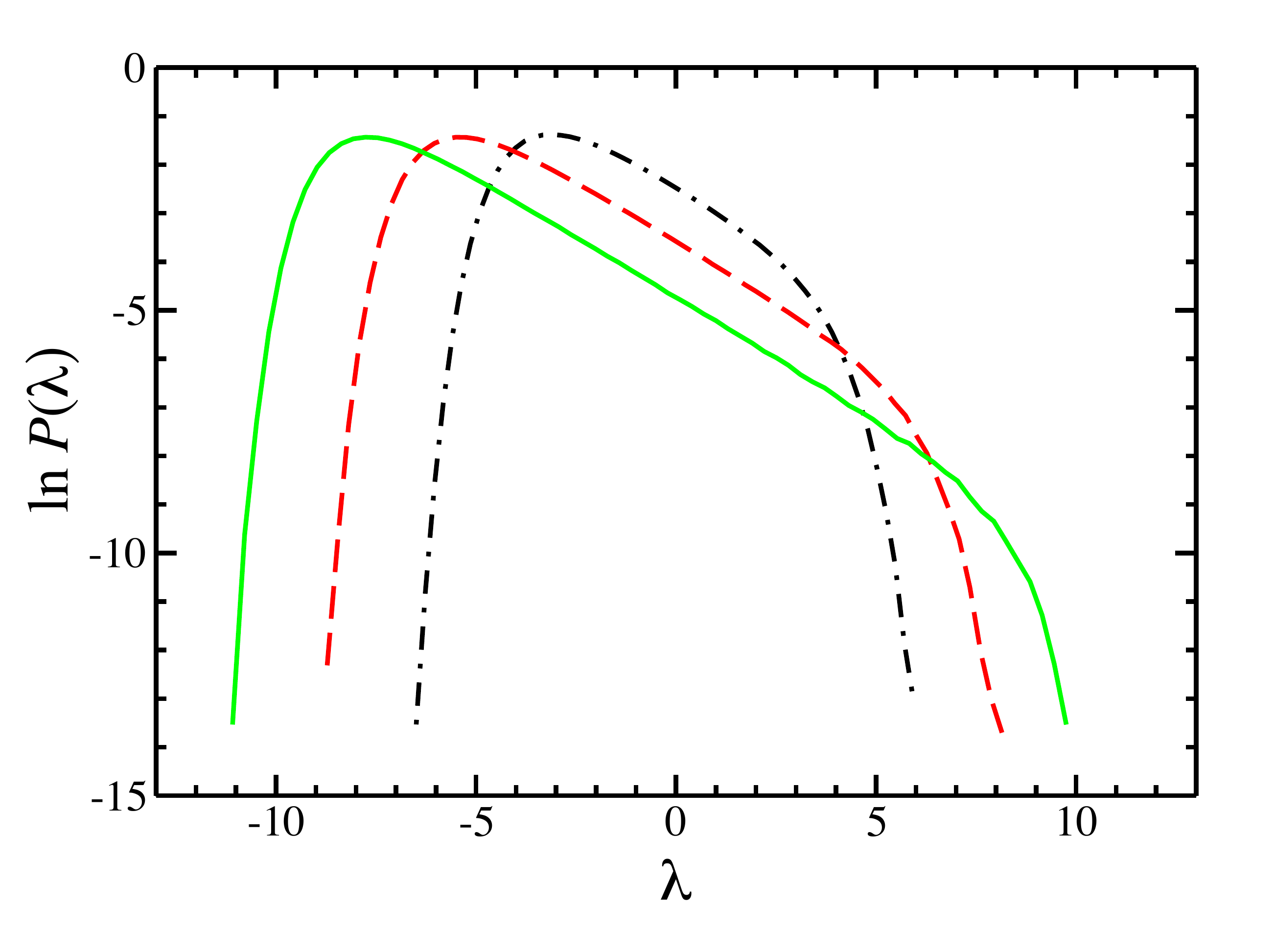}
\vspace{-0cm}
\end{center}
\caption{\label{fig:tailQ} (color online)
$\ln{P}(\lambda)$ for $\omega/\omega_p=0.5$, $C^g/C=0.01$,
$\sigma_S^2=0.01$, $\sigma_g^2=0$, and the quality factor
$Q=10^5,10^6,10^7$ (the dot-dashed, dashed, and solid curves,
respectively), corresponding to $\kappa\xi=4.6\times{10}^{-2},
4.6\times{10}^{-3},4.6\times{10}^{-4}$, respectively.}
\end{figure}

\begin{figure}
\begin{center}
\vspace{0cm}
\includegraphics[width=8cm]{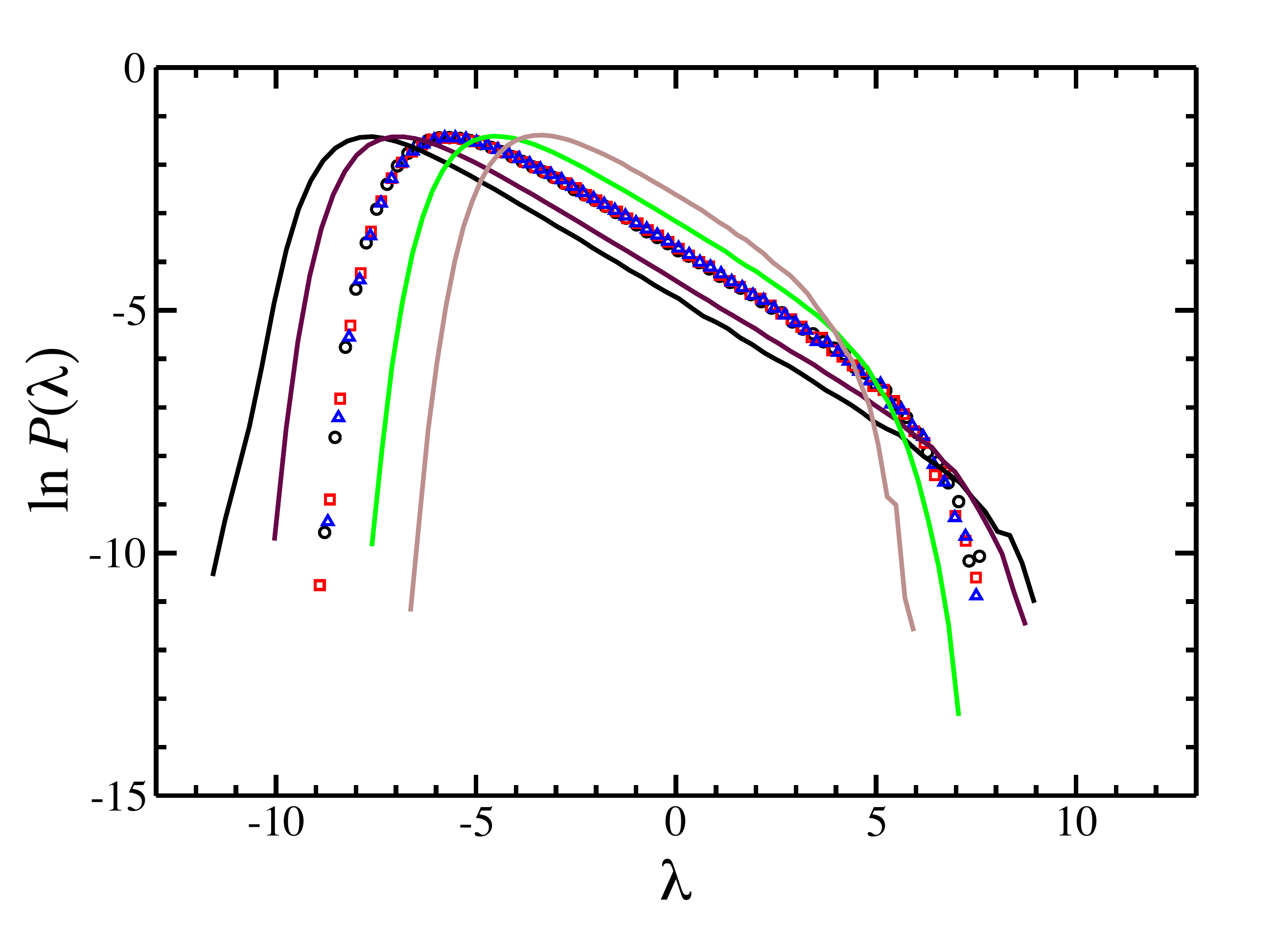}
\vspace{-0cm}
\end{center}
\caption{\label{fig:tailS} (color online)
$\ln{P}(\lambda)$ for $\omega/\omega_p=0.8$, $C^g/C=0.01$
and $Q=2\times{10}^6$.
The three sets of symbols represent the distributions for
$\sigma_S^2=0.01$, $\sigma_g^2=0$ (circles),
$\sigma_S^2=0.005$, $\sigma_g^2=0.005$ (squares),
and $\sigma_S^2=0$, $\sigma_g^2=0.01$ (triangles), all
corresponding to $\kappa\xi=3.3\times{10}^{-3}$.
They collapse to one curve, showing that the distribution is
sensitive only to the combination $\sigma_S^2+\sigma_g^2$.
The solid curves represent the distributions for $\sigma_g^2=0$
and $\sigma_S^2=0.05,\;0.03,\;0.003\;0.001$ (the curves with the
longer flat part corresponding to larger~$\sigma_S^2$), for which
the parameter $\kappa\xi=6.6\times{10}^{-4},\;1.1\times{10}^{-3},\;
1.1\times{10}^{-2},\;3.3\times{10}^{-2}$, respectively.}
\end{figure}

\begin{figure}
\begin{center}
\vspace{0cm}
\includegraphics[width=8cm]{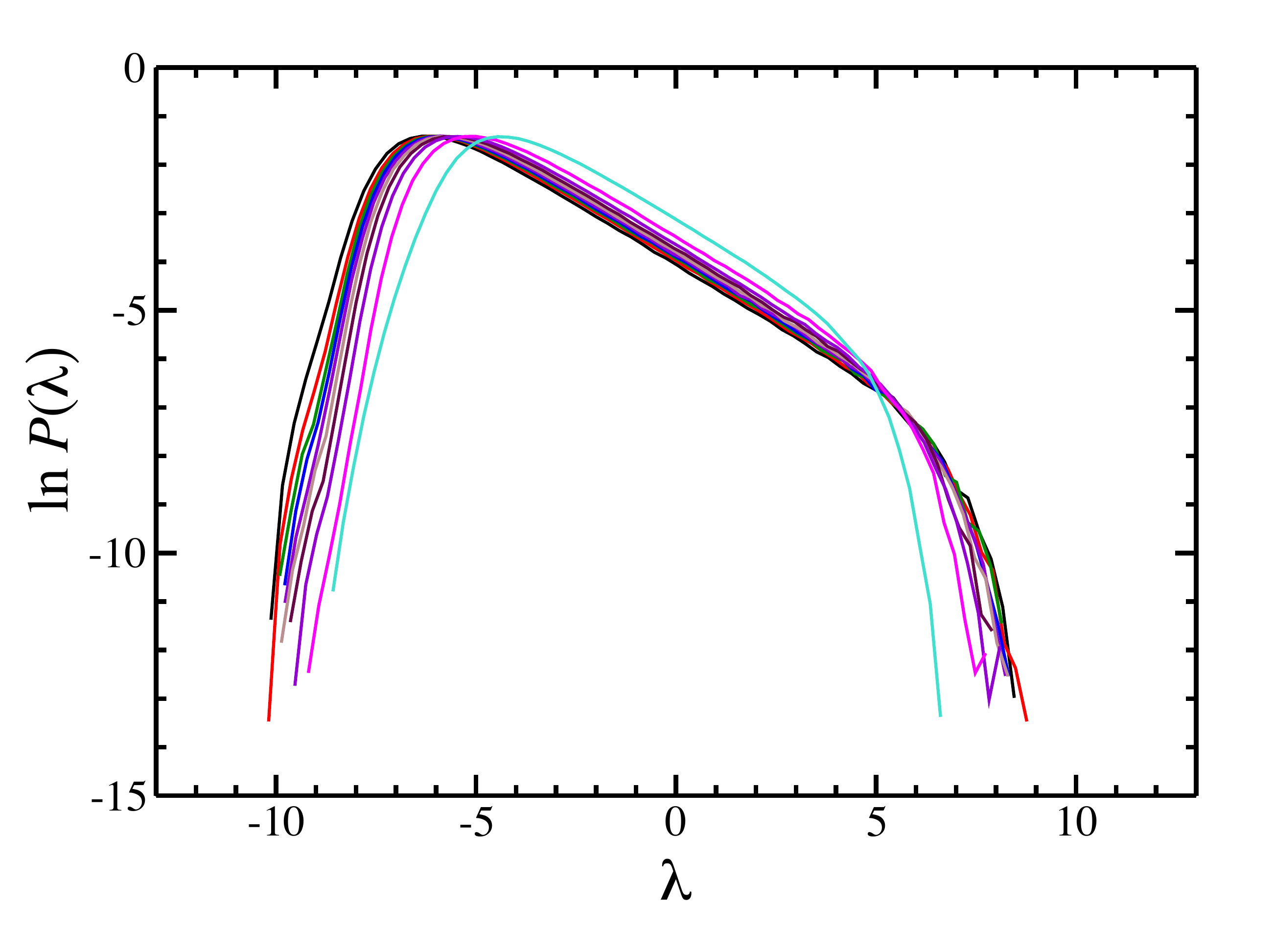}
\vspace{-0cm}
\end{center}
\caption{\label{fig:tailW} (color online)
$\ln{P}(\lambda)$ for $C^g/C=0.05$, $Q=10^5$, $\sigma_S^2=0.05$,
$\sigma_g^2=0$, and
$\omega/\omega_p=0.1,\,0.2,\,0.3,\,0.4,\,0.5,\,0.6,\,0.7,\,0.8,\,0.9,\,0.99$.
corresponding to $\kappa\xi=3.60\times{10}^{-3},\,3.65\times{10}^{-3},\,
3.75\times{10}^{-3},\,3.90\times{10}^{-3},\,4.12\times{10}^{-3},\,
4.46\times{10}^{-3},\,4.98\times{10}^{-3},\,5.90\times{10}^{-3},\,
7.99\times{10}^{-3},\,1.57\times{10}^{-2}$, respectively.
The curves with the longer flat part correspond to lower frequencies.
}
\end{figure}

\begin{figure}
\begin{center}
\vspace{0cm}
\includegraphics[width=8cm]{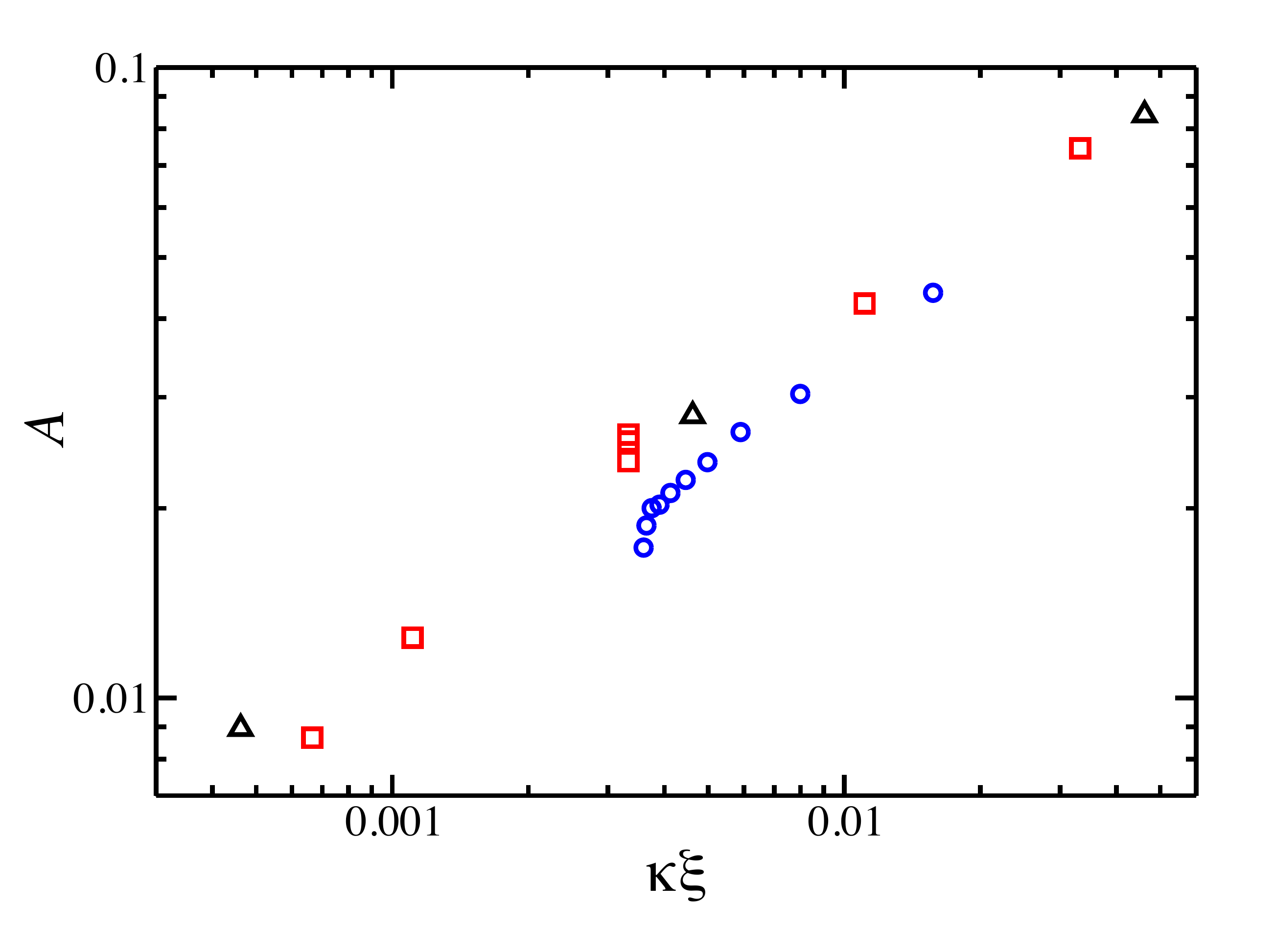}
\vspace{-0cm}
\end{center}
\caption{\label{fig:coeff} (color online)
The coefficient~$A$ in Eq.~(\ref{powerlaw=}) versus the parameter
$\kappa\xi$ for the distributions shown in
Fig.~\ref{fig:tailQ} (triangles),
Fig.~\ref{fig:tailS} (squares) and
Fig.~\ref{fig:tailW} (circles).
}
\end{figure}

For all curves $\ln{P}(\lambda)$ versus~$\lambda$ shown in
Figs.~\ref{fig:tailQ}, \ref{fig:tailS}, and \ref{fig:tailW} one can see a
flat part corresponding to a power-law tail, being more pronounced
for smaller $\kappa\xi$. For all curves the slope corresponds
to $\alpha-1\approx{0}.5$ (within a few percent), in agreement with
Eq.~(\ref{PZtail=}). The coefficient~$A$
in Eq.~(\ref{powerlaw=}), determined from the offset for all curves in
Figs.~\ref{fig:tailQ},\ref{fig:tailS},\ref{fig:tailW}, is plotted in
Fig.~\ref{fig:coeff} versus the parameter $\kappa\xi$. The points are
reasonably close to a straight line corresponding to
$A\propto\sqrt{\kappa\xi}$ with the coefficient between 0.3 and 0.5,
as stated in Eq.~(\ref{PZtail=}). However, we cannot exclude that the
deviations from the straight line are not just due to numerical reasons
(insufficient statistics, poor convergence, etc.) and signal the true
invalidity of single-parameter scaling in the strong fluctuation
regime.

\section{Conclusions}

In this paper, we analyzed the properties of the normal modes of a chain of Josephson
junctions in the superconducting regime, in the simultaneous presence of disorder and
absorption. We considered the limit where disorder and absorption can be treated
additively and computed the frequency shift and the localization length of the modes. We
also calculated the distribution of the frequency-dependent impedance of the chain. The
statistics depend on the parameter $\kappa \xi$, the ratio of the localization length and
the absorption length. If $\kappa \xi \gg 1$, the modes within one localization length
are much broadened by absorption and strongly overlap. This is the regime of weak
fluctuations; the distribution is Gaussian. In the opposite limit of little broadening,
the modes present within one localization length are well separated in frequency and
fluctuations are strong; the distribution has a power law tail.

The frequency-dependent impedance of Josephson junction chains can be probed
experimentally in principle, {\em e.g.}, by incorporating the chain in a resonator which
is capacitively coupled to a co-planar wave guide (CPW)~\cite{Masluk2012}. Microwave
transmission experiments on the CPW enable one to probe the small oscillation modes of
the chain directly. Alternatively, one could include a weaker junction with a small
Josephson energy (a so-called black-sheep junction~\cite{Manucharyan2012}) into the
chain, and measure its dc current-voltage characteristics. The low-voltage dc response
yields information about the ac impedance of the chain at frequencies $\sim 2eV/\hbar$,
as discussed in detail in Appendix~\ref{app:BS}.

Finally, Josephson junction chains have been predicted to constitute quantum phase-slip
(QPS) elements~\cite{Matveev2002}. Recent experiments have provided evidence for the
occurrence of QPS in Josephson junction
chains~\cite{Pop2010,Manucharyan2012,Haviland2013}. Phase-slip elements are of interest
for applications~\cite{Mooij2006}, {\em e.g.}, as qubits for quantum information
processing~\cite{Astafiev2012} and as a current standard for quantum
metrology~\cite{Guichard2010}. A typical quantum phase-slip event generally excites the
normal modes of the chain, therefore the QPS amplitude strongly depends on spectral
properties of the modes~\cite{Rastelli2012}. It would be interesting to calculate the QPS
amplitude for disordered Josephson chains. In the presence of disorder the phase-slip
process could be decoupled from part of the modes, a fact that might well result in an
enhancement of the phase-slip amplitude.

\section{Acknowledgements}
We thank G.~Crisan, A.~DiMarco, V.~Golovach, W.~Guichard, G.~Rastelli, and T.~Ziman for
useful discussions. We acknowledge financial support from the European network SOLID,
from Institut universitaire de France, and from the European Research Council (grant
``FrequJoc'' No.~306731).
\appendix

\section{Relation between impedance and reflection}
\label{app:reflection}

To define the amplitude reflection coefficient of a Josephson
junction chain, one should consider the system~(\ref{linsys=})
with all $I_n=0$,
replace the part of the chain to the left of the island $n=N_1$
by an effective impedance $Z_\infty$, and assume the part of
the chain on the right to be disorder-free and dissipationless.
Thus, the equation for
$n=N_1$ should be replaced by the effective boundary condition
\begin{equation}
\left(-\frac{1}{i\omega{L}}-i\omega{C}\right)
(V_{N_1}-V_{N_1+1})+\frac{V_{N_1}}{Z_\infty}=0,\quad
\end{equation}
and the solution of the system for $n\geq{N}_1$ should be sought
in the form $V_n=e^{-ik(n-N_1)}+re^{ik(n-N_1)}$, where the wave
vector~$k$ is related to the frequency $\omega$ by the dispersion
relation~(\ref{cleandisp=}) at $Q\to\infty$. This gives
\begin{equation}\label{reflexion=}
r=-\frac{1+i\omega{C}^gZ_\infty\displaystyle[\ep_\omega
+i\sqrt{1-(1-\ep_\omega)^2}]/(2\ep_\omega)}%
{1+i\omega{C}^gZ_\infty[\ep_\omega
-i\sqrt{1-(1-\ep_\omega)^2}]/(2\ep_\omega)},
\end{equation}
where $\ep_\omega\equiv(\omega^2LC^g/2)/(1-\omega^2LC)$.
If $Z_\infty$ is real, then $|r|=1$. Otherwise,
\begin{widetext}
\begin{equation}
1-|r|^2=\frac{2\sqrt{1-(1-\ep_\omega)^2}\,\omega{C}^g\Re{Z}_\infty}%
{\sqrt{1-(1-\ep_\omega)^2}\,\omega{C}^g\Re{Z}_\infty+\ep_\omega
-\ep_\omega\omega{C}^g\Im{Z}_\infty+|\omega{C}^gZ_\infty|^2/2}.
\end{equation}
\end{widetext}
The one-to-one correspondence between $Z_\infty$ and~$r$, expressed
by Eq.~(\ref{reflexion=}), implies that the numerical results of
Sec.~\ref{sec:Impedance} for the distribution of $\Re{Z}_\infty$
in the regime $\kappa\xi\ll{1}$ can be straightforwardly translated
into the distribution of $|r|^2$, which turns out to have a power-law
tail $\propto(1-|r|^2)^{-2.0}$, in agreement with
Ref.~\onlinecite{PradhanKumar}.

\section{Probability distribution of a sum of Lorentzians}
\label{app:distribution}

Instead of the probability distribution function, we first calculate
its Laplace transform called the characteristic function,
\begin{equation}\begin{split}
\mathcal{Q}(s)\equiv{}&{}\int\limits_0^\infty{e}^{-sx}P(x)\,dx=\\
={}&{}\int\limits_{-\Delta/2}^{\Delta/2}
\frac{d\omega_1}\Delta\ldots\frac{d\omega_{N_\xi}}\Delta\,
\exp\left(-\sum_{\alpha=1}^{N_\xi}
\frac{s\delta_\xi\gamma/\pi}{\omega_\alpha^2+\gamma^2}\right),
\end{split}\end{equation}
where we have shifted the integration variables by~$\omega$.
The $N_\xi$-fold integral is factorised into a product of
$N_\xi$~identical integrals. Recalling that
$N_\xi=\Delta/\delta_\xi$ we represent each such integral as
\begin{equation}\begin{split}
&\int\limits_{-N_\xi\delta_\xi/2}^{N_\xi\delta_\xi/2}
\frac{d\omega_1}{N_\xi\delta_\xi}\,
\exp\left(-\frac{s\delta_\xi\gamma/\pi}{\omega_1^2+\gamma^2}\right)=\\
&=1-\frac{\gamma}{N_\xi\delta_\xi}\int\limits_{-\infty}^\infty{d}y
\left\{1-\exp\left[-\frac{s\delta_\xi/(\pi\gamma)}%
{y^2+1}\right]\right\},
\end{split}\end{equation}
where we introduced the dimensionless variable $y=\omega_1/\gamma$
and used the fact that the exponential is different from unity only
limits of the last integral to infinity. Since this last integral no
longer depends on~$N_\xi$, the characteristic function can be
represented as
\begin{equation}
\mathcal{Q}(s)=\left(1-\frac{\mathcal{F}(s)}{N_\xi}\right)^{N_\xi}
\approx{e}^{-\mathcal{F}(s)},
\end{equation}
where we took the limit $N_\xi\to\infty$. The function
$\mathcal{F}(s)$ can be calculated exactly:
\begin{equation}\begin{split}
\mathcal{F}(s)&=\frac{\gamma}{\delta_\xi}\int\limits_{-\infty}^\infty{d}y
\int\limits_0^{s\delta_\xi/(\pi\gamma)}\frac{e^{-z/(y^2+1)}dz}{y^2+1}=\\
{}&{}=\frac{\pi\gamma}{\delta_\xi}
\int\limits_0^{s\delta_\xi/(\pi\gamma)}e^{-z/2}I_0(z/2)\,dz=\\
{}&{}=s
\exp\left(-\frac{s\delta_\xi}{2\pi\gamma}\right)
\left[I_0\!\left(\frac{s\delta_\xi}{2\pi\gamma}\right)
+I_1\!\left(\frac{s\delta_\xi}{2\pi\gamma}\right)\right],
\label{Fs=}
\end{split}\end{equation}
where $I_0,I_1$ are the modified Bessel functions.
This immediately gives us access to the moments of $P(x)$:
\begin{subequations}\begin{eqnarray}
&&\langle{1}\rangle=e^{-\mathcal{F}(0)}=1,\\
&&\langle{x}\rangle=\mathcal{F}'(0)=1,\\
&&\langle{x}^2\rangle-\langle{x}\rangle^2=-\mathcal{F}''(0)
=\frac{\delta_\xi}{2\pi\gamma}.
\end{eqnarray}\end{subequations}
The first two equations are trivial (recall that we have
defined~$x$ relative to the average value), but the last one
already tells us that the fluctuations become large when
$\delta_\xi/\gamma\gg{1}$.

To obtain the distribution function $P(x)$, we have to perform
the inverse Laplace transform of $e^{-\mathcal{F}(s)}$. Unable
to do it for the exact expression~(\ref{Fs=}), we use two
asymptotic expressions:
\begin{equation}
\mathcal{F}(s)=\left\{\begin{array}{ll}
s-s^2\delta_\xi/(4\pi\gamma),&s\delta_\xi/\gamma\ll{1},\\
\sqrt{4s\gamma/\delta_\xi},&s\delta_\xi/\gamma\gg{1}.
\end{array}\right.
\label{Fsasymp=}
\end{equation}
The most important values of~$s$ for the reconstruction of
$P(x)$ are those for which $\mathcal{F}(s)\sim{1}$, that is,
$s\sim\max\{1,\delta_\xi/\gamma\}$. Thus, the first expression
from Eq.~(\ref{Fsasymp=}) is good for the limit
$\delta_\xi\ll\gamma$, while the second one is good when
$\delta_\xi\gg\gamma$. One can check the that the inverse
Laplace transforms of $e^{-\mathcal{F}(s)}$ for the two
expressions are given by
\begin{subequations}\begin{eqnarray}
&&P(x)=\sqrt{\gamma/\delta_\xi}\,
e^{-(\pi\gamma/\delta_\xi)(x-1)^2},\delta_\xi\ll\gamma,
\label{PxGauss=}\\
&&P(x)=\frac{e^{-\gamma/(x\delta_\xi)}}%
{\sqrt{\pi(\delta_\xi/\gamma)x^3}},\quad
\delta_\xi\gg\gamma.\label{Pxtail=}
\end{eqnarray}\end{subequations}
For the first expression the check is straightforward, while
for the second one we have (rescaling $s\to{s}\gamma/\delta_\xi$)
\begin{equation}\begin{split}
&\int\limits_0^\infty\frac{e^{-1/y-sy}}{\sqrt{\pi{y}^3}}\,dy=
\quad(y=e^{2t})\\
&=\frac{2}{\sqrt\pi}\int\limits_{-\infty}^\infty
\exp\left(-t-e^{-2t}-se^{2t}\right)dt=
\quad\left(t\to{t}-\frac{\ln{s}}{4}\right)\\
&=\frac{2s^{1/4}}{\sqrt\pi}\int\limits_{-\infty}^\infty%
e^{-t-2\sqrt{s}\cosh{2}t}dt=\\
&=\frac{4s^{1/4}}{\sqrt\pi}\int\limits_0^\infty
e^{-2\sqrt{s}\cosh{2t}}\cosh{t}\,dt=
\quad(\sinh{t}=u)\\
&=\frac{4s^{1/4}}{\sqrt\pi}\int\limits_0^\infty
e^{-2\sqrt{s}(2u^2+1)}\,du=e^{-\sqrt{4s}}.
\end{split}\end{equation}

While Eq.~(\ref{PxGauss=}) is good enough and coincides with
Eq.~(\ref{Pxdllg=}),
the distribution in Eq.~(\ref{Pxtail=}) has all moments divergent,
as a consequence of the non-analyticity of
$\mathcal{F}(s)=\sqrt{4s\gamma/\delta_\xi}$ at $s\to{0}$. This
non-analyticity is an artefact of the asymptotic expression we
have used, which loses its validity at small
$s\sim\gamma/\delta_\xi$, while the exact $\mathcal{F}(x)$ from
Eq.~(\ref{Fs=}) is always analytic at $s\to{0}$. In terms of
$P(x)$, this means that the large-$x$ tail $P(x)\propto{x}^{-3/2}$
should be cut off at $x\sim\delta_\xi/\gamma$.

The cutoff can be easily obtained directly from the definition
of $P(x)$, Eq.~(\ref{Pxdef=}), by noting that large~$x$ corresponds
to $\omega$~being close to one of the Lorentzians. Neglecting the
probability of overlap of two Lorentzians, and noting that any of
the $N_\xi$ Lorentizans can be close to~$\omega$, we obtain
\begin{equation}\begin{split}
P(x){}&{}=N_\xi\int\limits_{-\infty}^\infty\frac{d\omega}\Delta\,
\delta\!\left(x-\frac{\delta_\xi\gamma/\pi}{\omega^2+\gamma^2}\right)
=\\
{}&{}=\frac{1}{\pi{x}^{3/2}\sqrt{\delta_\xi/(\pi\gamma)-x}},
\end{split}\end{equation}
valid for $x\gg\gamma/\delta_\xi$ (which is the typical value
of~$x$ between the Lorentzians).
The previous expression, Eq.~(\ref{Pxtail=}), is valid for
$0<x\ll\delta_\xi/\gamma$. Thus, the two expressions are both
valid in the wide region
$\gamma/\delta_\xi\ll{x}\ll\delta_\xi/\gamma$, where they both
reduce to $1/\sqrt{\pi(\delta_\xi/\gamma)x^3}$. Combining the
two, we arrive at Eq.~(\ref{Pxdggg=}).

\section{Damped harmonic oscillator}
\label{app:harmonic}

Consider the Hamiltonian equations for a damped harmonic
oscillator of the mass~$m$ and eigenfrequency $\omega_0$,
subject to an external force~$f(t)$:
\begin{subequations}\begin{eqnarray}
&&\frac{dp}{dt}=-m\omega_0^2x-2\gamma{p}+f(t),\\
&&\frac{dx}{dt}=\frac{p}m,
\end{eqnarray}\end{subequations}
$2\gamma$~being the damping rate. Assuming the force to be
monochromatic and looking for the solutions
$\propto{e}^{-i\omega{t}}$, we obtain the equation
\begin{equation}\label{oscsys=}
(\omega_0^2-2i\gamma\omega-\omega^2)\,p=-i\omega{f},
\end{equation}
which is the analog of Eq.~(\ref{veclinsys=}). The analog
of Eq.~(\ref{C12vw=}) is then
\begin{equation}
\left(\omega-\hat{\mathcal{A}}\right)
\left(\begin{array}{c} p \\ w \end{array}\right)
=\left(\begin{array}{c} if \\ 0 \end{array}\right)
\end{equation}
with the matrix $\hat{\mathcal{A}}$ given by
\begin{equation}
\hat{\mathcal{A}}=\left(\begin{array}{cc}
-2i\gamma & \omega_0 \\ \omega_0 & 0
\end{array}\right).
\end{equation}
The auxiliary variable $w$ is nothing but $w=-ix$.

The matix $\hat{\mathcal{A}}$ has two non-degenerate eigenvalues
\[
-i\gamma\pm\sqrt{\omega_0^2-\gamma^2},
\]
except for the special case $\gamma=\omega_0$. In this case there
is one doubly degenerate eigenvalue, and the matix $\hat{\mathcal{A}}$
has only one eigenvector. For $\gamma<\omega_0$ (weak damping),
the two eigenvalues can be denoted by $\omega_1$ and $-\omega_1^*$.
The normalization condition, analogous to Eq.~(\ref{Vnorm=}),
\begin{equation}
p_1^2\left(1+\frac{\omega_0^2}{\omega_1^2}\right)=1,
\end{equation}
determines the oscillator moblity $b=f/p$, which is analogous to
the impedance in Eq.~(\ref{ZV=}):
\begin{equation}
b(\omega)=\frac{ip_1^2}{\omega-\omega_1}
+\frac{i(p_1^*)^2}{\omega+\omega_1^*}
=\frac{i\omega}{\omega^2-2i\gamma\omega-\omega_0^2},
\end{equation}
which, of course, also follows directly from Eq.~(\ref{oscsys=}).

\section{Current-voltage characteristic of a black-sheep junction
coupled to a Josephson chain}
\label{app:BS}

\begin{figure}
\begin{center}
\vspace{0cm}
\includegraphics[width=8cm]{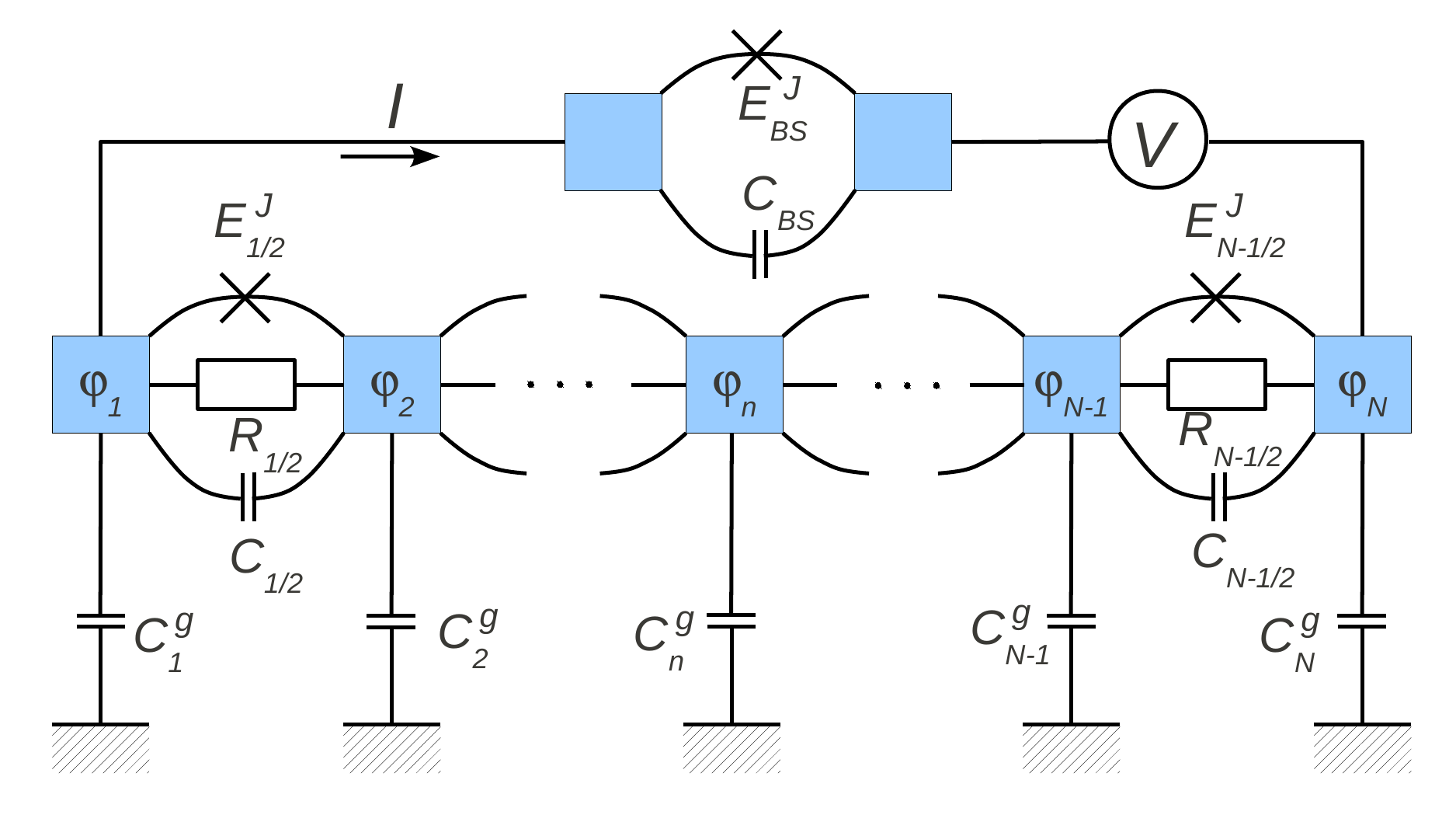}
\vspace{-0cm}
\end{center}
\caption{\label{fig:BS} (color online)
A schematic view of the black-sheep scheme of impedance measurement.
}
\end{figure}

We consider a voltage-biased circuit (bias voltage $V$) containing
a black-sheep (BS) junction with capacitance $C_{BS}$ and Josephson
coupling energy $E_{J,BS}$ in series with a Josephson junction chain
(Fig.~\ref{fig:BS}).
In the absence of quasiparticles, a small dc voltage bias~$V$ is
expected to induce a Cooper pair current $I(V)$. The necessary
dissipation is provided by the broadened modes of the Josephson
junction chain. For small $E_{J,BS}$, a perturbative calculation
yields~\cite{Ingold1992}
\begin{equation}
I(V) = \frac{\pi e E_{J,BS}^2}{\hbar}[P(2eV) - P(-2eV)]. \label{IV=}
\end{equation}
Here we defined the function $P(E)$ as
\begin{equation}
P(E) = \frac{1}{2 \pi \hbar}\int dt\, e^{J(t) + iEt/\hbar} \label{PE=},
\end{equation}
with
\begin{equation}\begin{split}
J(t) ={}&{} 8  \int \limits_ 0 ^\infty \frac{d\omega}{\omega}\,
\frac{\Re[Z_\mathrm{tot}(\omega)]}{R_K} \times \\
{}&{} \times \left\{[\cos (\omega t)-1]\coth\frac{\beta \hbar \omega}{2}
 -i \sin(\omega t)\right\}\label{Jt=},
\end{split}\end{equation}
where $R_K = h/e^2$. The chain is kept at the inverse temperature $\beta = 1/k_BT$; the
impedance $Z_\mathrm{tot}$ is the total impedance "seen" by the BS junction: a parallel
arrangement of the junction capacitance $C_{BS}$ and the impedance $Z(\omega)$ of the
chain. Hence,
\begin{equation}
\Re[Z_\textrm{tot}(\omega)] = \frac{\Re [Z(\omega)]}{|1 - i \omega C_{BS} Z(\omega)|^2}\
.\label{ReZtot}
\end{equation}

An interesting case is the so-called weak-coupling limit, where $J(t)$ remains small on
the relevant time scales, such that
\begin{equation}
\label{PE.exp}P(E) \simeq\frac{1}{2\pi\hbar}\int_{-\infty}^{+\infty}dt \,
e^{iEt/\hbar} \, [ 1+J(t)].
\end{equation}
This corresponds to the case where the BS junction exchanges at most one photon with the
chain. Indeed, the evaluation of the integral over time in (\ref{PE.exp}) gives
\begin{equation}\begin{split}
&P(E)\simeq\delta(E)+8\int_0^{+\infty}\frac{d\omega}{\omega} \,
\frac{\Re[Z_{\textrm{tot}}(\omega)]}{R_K}  \times \\
&\times\left[n_\omega\delta(E+\hbar\omega)+
(n_\omega+1)\delta(E-{\hbar}\omega)
\label{PEweak.one} -(2n_\omega+1)\delta(E)\right],
\end{split}\end{equation}
where $n_\omega=1/(e^{\beta\hbar\omega}-1)$ is the Bose-Einstein distribution function.
The first and the fourth terms represent elastic Cooper pair tunnelling in the BS junction
involving zero and one virtual photon, respectively. The second and third terms are
related to the process of absorption and emission of one real photon, respectively.

As can be seen from Eq.~(\ref{IV=}), the calculation of the current-voltage
characteristic involves the inelastic part of $P(E)$ only, given by
\begin{equation}
\label{PE.weak} P(E)\simeq \  8 \
\frac{\Re[Z_\textrm{tot}(E/\hbar)]}{R_K}\,\frac{1+n_{E/\hbar}}{E} .
\end{equation}
Substituting
Eq.~(\ref{PE.weak}) into Eq.~(\ref{IV=}) then yields the current-voltage characteristic
in the weak-coupling limit,
\begin{equation}
I(V)\simeq \frac{4\pi  E_{J,BS}^2}{\hbar V} \frac{\Re[Z_\textrm{tot}(2eV/\hbar)]}{R_K} \
.
\end{equation}
Note that the DC current at voltage $V$ directly probes the environmental impedance
$Z_\textrm{tot}$ at frequency $\omega = 2 eV/\hbar$. Specifically, at low frequencies,
the chain tends to become purely inductive, $\Re[Z(\omega \to 0)] \simeq N \omega^2
\langle L^2\rangle /R$ with $N$ the number of junctions, $\langle L^2\rangle$ the average
squared Josephson inductance of the chain and $R$ the junction resistance. Hence
$\Re[Z_\textrm{tot}(\omega \to 0)] \simeq N \omega^2 \langle L^2\rangle /R$. As a result,
at low voltage the current-voltage characteristic vanishes linearly with $V$. At high
frequencies, capacitive behavior takes over and for frequencies above the plasma
frequency the impedance tends tend to zero again. Hence the current also vanishes at high
voltages. Between zero and the plasma frequency, $\Re[Z_\textrm{tot}(\omega )]$ presents
a series of $N$ peaks, corresponding to the chain's modes. These will appear as current
peaks in the current-voltage characteristics.

\end{document}